\documentstyle[12pt]{article}

\makeatletter
\@addtoreset{equation}{section}
\makeatother

\newcommand{\bz}{{\hbox{\bf Z}}}
\newcommand{\br}{{\hbox{\bf R}}}
\newcommand{\bc}{{\hbox{\bf C}}}

\newcommand \ket[1]{\left\vert\, {#1} \, \right>}

\newcommand \qint[1]{\left[ {#1} \right]}

\newcommand \bosal[5]{\, \Phi  (
{#1};{#2},{#3} \, | \, {#4} ; {#5} )\,}
\newcommand \bosbl[5]{\, \phi  (
{#1};{#2},{#3} \, | \, {#4} ; { #5})\,}
\newcommand \boscl[5]{\, \chi  (
{#1};{#2},{#3} \, | \, {#4} ; { #5} )\,}
\newcommand \bosxl[5]{\, X  (
{#1};{#2},{#3} \, | \, {#4} ; { #5} )\,}
\newcommand \bosa[3]{\, \Phi  (
{#1} \, | \, {#2} ; { #3} )\,}
\newcommand \bosb[3]{\, \phi  (
{#1} \, | \, {#2} ; { #3} )\,}
\newcommand \bosc[3]{\, \chi  (
{#1} \, | \, {#2} ; { #3} )\,}
\newcommand \bosx[3]{\, X  (
{#1} \, | \, {#2} ; { #3} )\,}
\newcommand \pbosae[2]{\, \partial\Phi^{(\epsilon)}  (
 {#1} ; { #2} )\,}

\newcommand \aPhi[1]{a_{\Phi,{#1}}}
\newcommand \aphi[1]{a_{\phi,{#1}}}
\newcommand \achi[1]{a_{\chi,{#1}}}
\newcommand \aX[1]{a_{X,{#1}}}
\newcommand \Pup[1]{\Phi^{\lambda_+}_{\lambda,+}({#1})}
\newcommand \Pdn[1]{\Phi^{\lambda}_{\lambda_+,+}({#1})}
\newcommand \Pdnm[1]{\Phi^{\lambda}_{\lambda_+,-}({#1})}
\newcommand \Pupm[1]{\Phi^{\lambda_+}_{\lambda,-}({#1})}
\newcommand \Pp[1]{\Phi_{1,0}({#1})}

\newcommand \diff[2]{{~}_{\scriptstyle {#1}}
\displaystyle \partial_{\scriptstyle {#2}} \,}
\newcommand \fra[2]{\displaystyle\frac{#1}{#2}}
\newcommand \tfra[2]{\textstyle\frac{#1}{#2}}
\newcommand \pGam[1]{\Gamma_p({#1})}
\newcommand \Fp[1]{ F_p({#1})}
\newcommand \cN[1]{{\cal N}_{{#1}}}
\newcommand \Prod[1]{\prod_{{#1}}}

\newcommand{\inp}[2]{({#1} ; {#2} )_{\infty}}
\newcommand{\indp}[3]{({#1} ; {#2},{#3} )_{\infty}}
\newcommand{\intp}[4]{({#1} ; {#2},{#3},{#4} )_{\infty}}
\newcommand{\jintinf}[1]{\displaystyle \int_{0}^{c \infty} d_{p} t_{#1}\;}
\newcommand{\jint}[2]{\displaystyle \int_{0}^{{#1}} d_{p} t_{{#2}}\;}
\newcommand{\jintd}[3]{\displaystyle \int_{#1}^{{#2}\infty} d_{p} t_{{#3}}\;}
\newcommand{\intb}[3]{\displaystyle \int_{0}^{1} d_{p} t  t^{\scriptstyle{#1}}
\displaystyle\frac{(pt;p)_\infty({#2}zt;p)_\infty}
            {({#3}t;p)_\infty(q^{-2}zt;p)_\infty}\;}
\newcommand{\defintb}[3]{\displaystyle \int_{0}^{1} d_{p} t
            t^{\scriptstyle{#1}}
            \displaystyle\frac{(pt;p)_\infty(p^{#2}zt;p)_\infty}
            {(p^{#3}t;p)_\infty(zt;p)_\infty}\;}
\newcommand{\intbd}[3]{\displaystyle \int_{p}^{\infty} d_{p} t
            t^{\scriptstyle{#1}}
            \displaystyle\frac{(p/t;p)_\infty({#2}z/t;p)_\infty}
            {({#3}/t;p)_\infty(q^{-2}z/t;p)_\infty}\;}
\newcommand{\hh}{\rule[-5mm]{0mm}{10mm}}

\newcommand{\jdn}{J^{-}_{-1}}
\newcommand{\jsup}{S_{1}}
\newcommand{\jsdn}{S_{-1}}
\newcommand{\jpup}{J^{+}_{1}}
\newcommand{\jpdn}{J^{+}_{-1}}

\font\germ=eufm10 scaled \magstep1
\def\goth#1{\hbox{\germ #1}}

\def\slth{\widehat{\goth{sl}}_2\hskip 1pt}
\def\uq{U_q(\widehat{\goth{g}})}

\def\uqa{U_q\bigl(\slth\bigr)}

\def\uqap{U'_q\bigl(\slth\bigr)}
\def\cLl{{\cal L}_{\lambda}}
\def\cLm{{\cal L}_{\mu}}
\def\wei{W^{\epsilon}_i}
\def\weip{W^{\epsilon}_{i'}}
\def\twei{\tilde W^{\epsilon}_{i}}
\def\tweip{\tilde W^{\epsilon}_{i'}}
\def\tdj{T^{\delta}_{j}}
\def\tdjp{T^{\delta}_{j'}}
\def\urh{U^{\rho}_{h}}
\def\urhp{U^{\rho}_{h'}}

\def\Tr{{\rm Tr}}
\def\VEp{\varepsilon}

\begin{document}

\begin{flushright}
YITP/K-1074 \\
May 1994
\end{flushright}
\vspace{24pt}
\begin{center}
\begin{Large}
{Free Field Representation of Quantum Affine Algebra $\uqa$} \par
\vskip 3mm
{and Form Factors in Higher Spin XXZ Model}
\end{Large}

\vspace{56pt}
Hitoshi KONNO\raisebox{2mm}{$\star$}

\vspace{6pt}
{\it Yukawa Institute for Theoretical Physics,} \\
{\it Kyoto University, Kyoto 606-01, Japan}

\vspace{58pt}

{ABSTRACT}
\end{center}

\vspace{30pt}
We consider the spin  $k/2$ XXZ model in the antiferomagnetic regime
using the free field realization of the quantum affine algebra $\uqa$
of level $k$. We give
a free field realization of the type II $q$-vertex operator,
which describes creation and
annihilation of physical particles in the model.
By taking a trace of the type I and the type II $q$-vertex operators
over the irreducible highest weight representation of
$\uqa$, we also derive an integral formula for form factors in this model.
Investigating the structure of
poles, we obtain a residue formula for
form factors, which is
a lattice analog of the higher spin extension of the
Smirnov's formula in the massive integrable quantum field theory.
This result as well as the quantum deformation of the
Knizhnik-Zamolodchikov equation for  form factors
shows a deep connection in
the mathematical structure of the integrable lattice models and the
massive integrable quantum field theory.
\vspace{24pt}

\vfill
\hrule

\vskip 3mm
\begin{small}

\noindent\raisebox{2mm}{$\star$} Soryushi-Shogakukai Fellow

\end{small}

\newpage

\section{Introduction}
    Recently, in the attempts to clarify the mathematical structure of the
massive
integrable quantum  field theory (MIQFT),
quantum deformation of the affine Lie algebra is recognized as a
key symmetry to the integrability of the theory
\cite{Sma,BL}. Especially,
Smirnov has argued
that his equation for form factors in $SU(2)$ invariant Thirring model is
noting but a quantum  deformation of the  Knizhnik-Zamolodchikov (KZ)
equation and that the
Yangian double, a kind of quantum deformation of the affine Lie
algebra $\slth$ with zero central charge, is the key symmetry
to the equation. ( See also the third paper in Ref.\cite{BL})\par
     However, in the Smirnov's work, the solutions of the quantum
deformed KZ equation,
i.e form factors,
are not well understood in the
language of the representation theory of the Yangian.\par
     On the other hand,
Frenkel and Reshetikhin\cite{FR}
derived a quantum deformation of the KZ equation based on the quantum
affine algebra $\uq$ and related it to integrable lattice models.
Further, using the representation
of the quantum affine algebra $\uqa$,  people in Kyoto school  reformulated
the XXZ model in the
antiferromagnetic regime\cite{DFJMN,JMMN}
  and extended it to the higher spin cases\cite{IIJMNT}.
There, in contrast to the Smirnov's works,
the form factors are well defined in terms of vertex operators and
highest weight modules. \par
     An important observation made in \cite{DFJMN,IIJMNT} is that
the quantum KZ equation  for the
form factors in the XXZ model  coincides with Smirnov's deformed KZ equation
in a certain  scaling limit.
This suggests a deep connection in the mathematical structures
between  the XXZ model and the
$SU(2)$ invariant Thirring model, especially
 a possibility of understanding the
Smirnov's results in terms of the representation theory of quantum affine
algebra.\par
     The main purpose of this paper is to inquire further into this
 connection.
More precisely, we show that the form factor defined by Idzumi
et al.\cite{IIJMNT} based
on the representation theory of the quantum affine algebra $\uqa$ satisfies
a lattice analog of the axioms for form
factors in MIQFT invented by Smirnov\cite{Smb}.\par

     The Smirnov's axioms are summarized as follows.

\noindent{Axiom 1}\quad  Form factors
$f(\beta_1,..,\beta_n)_{\epsilon_1,..\epsilon_n}$ have the S-matrix symmetry
\begin{eqnarray}
\lefteqn{f(\beta_1,..,\beta_{i}\beta_{i+1},.
.,\beta_n)_{\epsilon_1,..,\epsilon_{i}\epsilon_{i+1},..,\epsilon_n}
S^{\epsilon'_i\epsilon'_{i+1}}_{\epsilon_i\epsilon_{i+1}}(\beta_i-\beta_{i+1})
} \nonumber \\
&& \qquad =f(\beta_1,..,\beta_{i+1}\beta_{i},.
.,\beta_n)_{\epsilon_1,..,\epsilon'_{i+1}\epsilon'_{i},..,\epsilon_n}
\label{axiom1}
\end {eqnarray}

\noindent{Axiom 2}\quad  Form factors
$f(\beta_1,..,\beta_n)_{\epsilon_1,..\epsilon_n}$ satisfy the
difference equation
\begin{equation}
f(\beta_1,..,\beta_n+2\pi i)_{\epsilon_1,.,\epsilon_n}
=f(\beta_n,\beta_1,..,\beta_{n-1})_{\epsilon_n,\epsilon_{1},..,\epsilon_{n-1}}
\label{axiom2}
\end{equation}

\noindent{Axiom 3}\quad Form factors
$f(\beta_1,..,\beta_n)_{\epsilon_1,..\epsilon_n}$ have  simple poles at
$\beta_i=\beta_j+\pi i,\quad i>j$. The  corresponding
residues at  $\beta_n=\beta_{n-1}+\pi i$  are given by
\par

\begin{eqnarray}
\lefteqn{2\pi i\ {{\rm res}}
f(\beta_1,..,\beta_n)_{\epsilon_1,..,\epsilon_n}
} \nonumber \\
&&=f(\beta_1,..,\beta_{n-2})_{\epsilon'_1,..,\epsilon'_{n-2}}
\delta_{\epsilon_n,-\epsilon_{n-1}}
 \nonumber \\
&&\times \biggl(\delta^{\epsilon'_1}_{\epsilon_1}\cdots
\delta^{\epsilon_{n-2}'}_{\epsilon_{n-2}}
-S^{\epsilon'_{n-1}\epsilon'_{1}}_{\tau_1\epsilon_{1}}(\beta_{n-1}-\beta_{1})
\cdots S^{\tau_{n-3}\epsilon'_{n-2}}_{\epsilon_{n-1}\epsilon_{n-2}}
(\beta_{n-1}-\beta_{n-2})\biggr).
\label{axiom3}
\end{eqnarray}

     These axioms define the  matrix Riemann-Hilbert problem, which provides
complete imfomations to derive  integral formulae for  form factors
of local operators in various MIQFTs\cite{Smb}. Especially, one should note
that the deformed KZ equation for form factors
is derived from the first and the second axioms.

      In  Refs.\cite{DFJMN,JMMN,IIJMNT}, the correlation functions and the
form factors in the XXZ model
were reconstructed  as a trace of the product of some $q$-vertex operators
over an irreducible highest weight representation (IHWR) of
$\uqa$. As a result, the
lattice analogs of the first and the second axioms are understood as
an $R$-matrix symmetry of the $q$-vertex operators and the cyclic property of
trace.
We here consider the third axiom in the same spirit.
Namely, we verify the lattice analog of the
third axiom in terms of the $q$-vertex operators.
In  the case of the spin 1/2 XXZ model, the same subject was  discussed
by Pakuliak\cite{Pa}. However, his analysis is based on the result restricted
to the spin 1/2 case. In this paper, we consider more general
situations containing the higher spin extensions.\par
\par
      We also discuss an explicit calculation of form factors in the spin
$k/2$ XXZ model. Even though this is not necessary to verify the third axiom,
this subject is important in the physical application.
We explicitly evaluate a trace of the $q$-vertex operators
over IHWR of $\uqa$ and obtain an integral formula for form factors.\par
      For these purposes, we make a full use of  the
free field realization of the quantum affine algebra $\uqa$
of level $k ( \geq 1)$\cite{Sh,Ma,ABG,Koa}. This realization is a
higher level extension of the level one realization
obtained in Ref.\cite{FJ,JMMN}. The characteristic
feature of our realization is the existence of screening operators and
their Jackson integrals ( screening charges ) in the expression of the
$q$-vertex operators.
For the purpose of physical application,
it is required to determine the cycles for the
Jackson integrals.  We discuss the determination of the
cycles, which leads to the $R$-matrix
symmetry of the $q$-vertex operators.
\par

     Furthermore, as in the free field representation of the affine
Lie algebras\cite{Fe,BF}, the
Fock space representation ( quantum deformation of the Wakimoto module\cite{Wa}
 )
of $\uqa$ is, in general, reducible.
This makes no sense  in the simple minded evaluation of traces over
the $q$-Wakimoto modules.
In Ref.\cite{Koa}, extending the  Felder's BRST
analysis of the affine Lie algebra $\slth$ to the case $\uqa$, we discussed a
resolution of IHWR of $\uqa$ and gave a formula, which allows us to evaluate
a trace over the IHWR in terms of free fields.
Our trace calculation is totally based  on  this result.
\par

     This paper is organized as follows. In the next section,
we briefly review the free field realization of
$\uqa$
and the resolution of IHWR based on the BRST-Felder cohomology analysis.
The prescriptions of evaluating trace over IHWR are also presented.\par
     In Sec.3, we derive a free field realization of the
type II $q$-vertex operator of $\uqa$ of level $k$,
which remains to be done in Refs.\cite{Sh,Ma,ABG,Koa}.
This vertex plays a role of an annihilation operator
(it's anti-pode dual plays a creation operator) of physical particle
in the  spin $k/2$ XXZ model.
We determine suitable cycles for the Jackson integral associated with
the screening charges by demanding  commutativity with all the currents
realizing $\uqa$. Calculating two point functions
of the type II $q$-vertex operators, we show
that these cycles imply
the $R$-matrix symmetry of the type II $q$-vertex operators.\par
        In Sec.4, we make an explicit evaluation of the
trace of the $q$-vertex operators over IHWR for  $\uqa$. We give a
general  integral formula for form factors.

        In Sec.5,
we investigate a pole structure of form factors.
It is shown that only a simple pole appears, when two spectral
parameters close each other.  Evaluating residues, we obtain a higher spin
extension of a lattice analog of the Smirnov's residue formula
for MIQFT ( the third axiom (\ref{axiom3})).

The final section is devoted to a discussion
 of  physical implication of the results.
\par

\section{Free Field Realization of $\uqa$ }

We here briefly review the free field realization of the
quantum affine algebra $\uqa$ and its Fock space
representation\cite{Sh,Ma,ABG,Koa}.

\vskip 2mm
\noindent{\bf 2.1 Drinfeld realization of $\uqap$}\quad

The quantum affine algebra $\uqa$ consists of
the algebra $\uqap$ and the grading operator $L_0$.
Let us first consider $\uqap$.
We follows the Drinfeld realization of $\uqap$\cite{Dr},
 which is
generated by the letters
$\{ J^{\pm}_{n} |n \in \bz \}$, $\{ J^{3}_{n} |n \in \bz_{\neq 0} \}$,
$\gamma^{\pm1/2}$ and $K$ under the
relations
\begin{equation}
\begin{array}{l}
\gamma^{\pm1/2} \in \mbox{ the center of the algebra}, \hh \\
\qint{J^{3}_{n},J^{3}_{m}} =
\delta_{n+m,0} \fra{1}{n}\qint{2n}
\frac{\gamma^{k}-\gamma^{-k}}{q-q^{-1}} , \hh \\
\qint{J^{3}_{n},K} = 0 , \hh \\
KJ^{\pm}_{n}K^{-1} = q^{\pm2}J^{\pm}_{n} , \hh \\
\qint{J^{3}_{n},J^{\pm}_{m}} =
\pm \fra{1}{n}\qint{2n} \gamma^{\mp|n|/2}J^{\pm}_{n+m} , \hh \\
J^{\pm}_{n+1}J^{\pm}_{m}-q^{\pm2}J^{\pm}_{m}J^{\pm}_{n+1} =
q^{\pm2}J^{\pm}_{n}J^{\pm}_{m+1}-J^{\pm}_{m+1}J^{\pm}_{n} , \hh \\
\qint{J^{+}_{n},J^{-}_{m}} =
\fra{1}{q-q^{-1}}(\gamma^{(n-m)/2}\psi_{n+m} -
\gamma^{(m-n)/2}\varphi_{n+m}). \hh
\end{array}
\label{Dreal}
\end{equation}
Here $\{ \psi_{r}, \varphi_{s} |r, s \in \bz \}$ are related to
$\{ J^{3}_{l} | l \in \bz_{\neq 0} \}$ as follows.
\begin{equation}
\begin{array}{l}
\displaystyle \sum_{n \in \bz} \psi_{n} z^{-n} = K
\exp \left\{ (q-q^{-1}) \sum^{\infty}_{k=1} J^{3}_{k}z^{-k} \right\} , \hh \\
\displaystyle \sum_{n \in \bz} \varphi_{n} z^{-n} = K^{-1}
\exp \left\{-(q-q^{-1}) \sum^{\infty}_{k=1} J^{3}_{-k}z^{k} \right\} . \hh
\end{array}
\end{equation}
In (\ref{Dreal}), we used the notation
$$
\qint{m} = \fra{q^{m}-q^{-m}}{q-q^{-1}}.
\label{Defqint}
$$

The standard Chevalley generators $\{ e_{i},f_{i},t_{i} \}$ are given
by the following identification
\begin{equation}
t_{0} = \gamma K^{-1}, \;\; t_{1} = K, \;\;
e_{1} = J^{+}_{0}, \;\; f_{1} = J^{-}_{0}, \;\;
e_{0} t_{1} = J^{-}_{1}, \;\; t^{-1}_{1} f_{0} = J^{+}_{-1} .
\end{equation}

Let us introduce three free bosonic fields
$\Phi, \phi$ and $\chi$
carrying parameters $L,M,N \in \bz_{>0}$, $\alpha \in \br$ defined
by\cite{Sh}
\begin{equation}
\begin{array}{rl}
\bosxl{L}{M}{N}{z}{\alpha} = &
- \displaystyle \sum_{n \neq 0}
\fra{ \qint{Ln} \aX{n}}
{ \qint{Mn}\qint{Nn}} z^{-n}q^{|n|\alpha}
+ \fra{ L\tilde\aX{0}}{ MN} \log z
+ \fra{ LQ_{X}}{ MN},
\end{array}
\end{equation}
for $X=\Phi, \phi$ or  $\chi$.
 We also often use the notation
\begin{equation}
\begin{array}{rl}
\bosx{N}{z}{\alpha} =&
\bosxl{L}{L}{N}{z}{\alpha} \\
=&  - \displaystyle \sum_{n \neq 0}
\fra{\aX{n}}
{ \qint{Nn}} z^{-n}q^{|n|\alpha}
+ \fra{ \tilde\aX{0}}{ N} \log z
+ \fra{ Q_{X}}{ N}.
\end{array}
\end{equation}

The fields are quantized by the commutation relations
\begin{equation}
\begin{array}{ll}
\qint{\aPhi{n},\aPhi{m}} = \delta_{n+m,0}
\fra{ \qint{2n}\qint{(k+2)n}}{ n}, &
\qint{\tilde{a}_{\Phi,0},Q_{\Phi}}=2(k+2), \hh \\
\qint{\aphi{n},\aphi{m}} = - \delta_{n+m,0}
\fra{ \qint{2n}\qint{2n}}{ n}, &
\qint{\tilde\aphi{0},Q_{\phi}}=-4, \hh \\
\qint{\achi{n},\achi{m}} = \delta_{n+m,0}
\fra{ \qint{2n}\qint{2n}}{ n}, &
\qint{\tilde\achi{0},Q_{\chi}}=4, \hh
\end{array}
\label{halge}
\end{equation}
where
$$
\begin{array}{c}
\tilde\aX{0}=\fra{ q-q^{-1}}{ 2\log q} \aX{0} , \hh
\end{array}
$$
for $X=\Phi, \phi$ and  $\chi$, and others commute.

By means of the fields $\Phi, \phi$ and $\chi$, we realize the generators
$J^{3}(z), J^{\pm}(z)$, $K$ and $\gamma$ in $\uqap$
as follows.
\begin{equation}
\begin{array}{rl}
J^{3}(z) =& \diff{k+2}{z} \bosa{k+2}{q^{-2}z}{-1}
+\diff{2}{z} \bosb{2}{q^{-k-2}z}{-{\tfra{k+2}{2}}} , \hh \\
J^{+}(z) =&
-:\left[\diff{1}{z} \exp \left\{ -\bosc{2}{q^{-k-2}z}{0} \right\} \right]
\times \exp
\left\{ -\bosb{2}{q^{-k-2}z}{1} \right\}: , \hh \\
J^{-}(z) =&
:\Bigl[ \diff{k+2}{z} \exp \Bigl\{
\bosa{k+2}{q^{-2}z}{-{\tfra{k+2}{2}}}
+\bosb{2}{q^{-k-2}z}{- 1}  \hh \\
& \hspace{2.5cm}
+\boscl{k+1}{2}{k+2}{q^{-k-2}z}{0} \Bigr\}\Bigr] \hh \\
& \times \exp \left\{-\bosa{k+2}{q^{-2}z}{\tfra{k+2}{2}}
+\boscl{1}{2}{k+2}{q^{-k-2}z}{0} \right\}: , \hh
\end{array}
\label{Jdef}
\end{equation}
\begin{equation}
   K= q^{\tilde\aPhi{0}+\tilde\aphi{0}}, \qquad \gamma = q^{k}.
  \label{Kdef}
\end{equation}
Here we defined the $q$-difference operator with parameter $n \in \bz_{>0}$ by
$$
\diff{n}{z} f(z) \equiv \fra{ f(q^{n}z) - f(q^{-n}z)}
{ (q-q^{-1})z}.
$$
The  mode expansions of the currents $J^3(z)$ and $J^{\pm}(z)$
should be set as follows.

\begin{eqnarray}
\sum_{n \in \bz} J^{3}_{n} z^{-n-1} &=& J^{3}(z) , \\
\sum_{n \in \bz} J^{\pm}_{n} z^{-n-1}&=& J^{\pm}(z).
\label{mode}
\end{eqnarray}

       Finally, the grading operator $L_0$
is realized as follows.

\begin{eqnarray}
     L_0&=&L_0^{\Phi}+L_0^{\phi}+L_0^{\chi} \\
     L_0^{\Phi}&=&\fra{\tilde\aPhi{0}(\tilde\aPhi{0}+2)}{4(k+2)}+
        \sum_{n\geq 1}\fra{n^2}{\qint{2n}\qint{(k+2)n}}\aPhi{-n}\aPhi{n},
       \\
     L_0^{\phi}&=&-\fra{\tilde\aphi{0}(\tilde\aphi{0}-2)}{8}-
        \sum_{n\geq 1}\fra{n^2}{{\qint{2n}}^2}\aphi{-n}\aphi{n}, \\
     L_0^{\chi}&=&\fra{\tilde\achi{0}(\tilde\achi{0}+2)}{8}+
        \sum_{n\geq 1}\fra{n^2}{{\qint{2n}}^2}\achi{-n}\achi{n}.
\label{grade}
\end{eqnarray}

\vskip 2mm
\noindent{\bf 2.2 Finite Dimensional $\uqa$ module } \quad

    Let   $V^{(l)}$ be
the $(l+1)-$dimensional $\uqa$ module with basis $v^{(l)}_m$, $m=0,1,2..,l$:
\begin{eqnarray}
e_1v^{(l)}_m &=&[m]v^{(l)}_{m-1},\qquad
f_1v^{(l)}_m =[l-m]v^{(l)}_{m+1},\qquad
t_1v^{(l)}_m =q^{l-2m}v^{(l)}_{m},\nonumber\\
e_0&=&f_1,\qquad f_0=e_1,\qquad t_0=t_1^{-1}\qquad  {\rm on}\quad V^{(l)},
\label{finitem}
\end{eqnarray}
where $v^{(l)}_m=0$ if $m<0$ or $m>l$.
In the case $l=1,$ we often use the notations
$v^{(1)}_0=v_+$ and $v^{(1)}_1=v_-$.

    Let $V^{(l)}_z$ be  the affinization of $V^{(l)}$.
The action of the Drinfeld generators on $V^{(l)}_z$ is given by\cite{Sh}
\begin{eqnarray}
\gamma^{\pm1/2}v^{(l)}_m &=&v^{(l)}_{m},\nonumber \\
Kv^{(l)}_m &=&q^{l-2m}v^{(l)}_{m},\nonumber \\
J^+_nv^{(l)}_m& =&z^nq^{n(l-2m+2)}[l-m+1]v^{(l)}_{m-1},\label{dfinitem} \\
J^-_nv^{(l)}_m &=&z^nq^{n(l-2m)}[m+1]v^{(l)}_{m+1},\nonumber \\
J^3_nv^{(l)}_m &=&\fra{z^n}{n}\Bigl\{[nl]-q^{n(l+1-m)}(q^n+q^{-n})[nm]\Bigr\}
                    v^{(l)}_{m}.\nonumber
\end{eqnarray}

\vskip 2mm
\noindent{\bf 2.3 $q$-Wakimoto module } \quad

The quantum deformation of the
Wakimoto module is defined as a certain restriction of the
Fock module of the three bosons $\Phi, \phi$ and $\chi$\cite{Sh,Ma,Koa}.
Let us briefly describe this restriction.\par

    Let us begin  with the following Fock module.
\begin{equation}
    F_{l,s,t}=\left\{ \prod_{n>0}\aPhi{-n}
                       \prod_{n'>0}\aphi{-n'}
                        \prod_{n''>0}\achi{-n''}|l;s,t>\right\}
\end{equation}
\begin{equation}
   |l;s,t>=\exp\left\{l\fra{Q_{\Phi}}{2(k+2)}
                       +s\fra{Q_{\phi}}{2}
                        +t\fra{Q_{\chi}}{2} \right\}|0>.
\end{equation}
Here $|0>$ is the vacuum state of the Heisenberg algebra (\ref{halge}).

     Let $\lambda_l, l=0,1,2,..,k$ be the dominant integral weights of level
$k$;
$\lambda_l=(k-l)\Lambda_0+l\Lambda_1$ where $\Lambda_0$ and  $\Lambda_1$ are
the fundamental weights. One can show that  the states
$\ket{\lambda_l}\equiv|l;0,0>$ give rise to
 the highest weight states satisfying
$t_1\ket{\lambda_l}=q^l\ket{\lambda_l},\
t_0\ket{\lambda_l}=q^{k-l}\ket{\lambda_l},\
e_i\ket{\lambda_l}=0, i=0,1$ and
$q^{L_0}\ket{\lambda_l}=q^{\Delta_{\lambda_l}}\ket{\lambda_l}$,
where $\Delta_{\lambda_l}=l(l+2)/4(k+2)$.

    Let us first fix the picture of the Fock module $F_{l,s,t}$\cite{FMS}.
This is carried out
by projecting $F_{l,s,t}$ onto the sector,
on which  the eigen value of the operator $\tilde \aphi{0}+\tilde \achi{0}$ is
zero. Then we have the picture fixed Fock module $F_{l,s,s}$. The
highest weight $\uqa$ module $V(\lambda_l)$ of weight $\lambda_l$
is  embeded
in the Fock module
$F_l\equiv \oplus_{s\in \bf Z}F_{l,s,s}$ \cite{Sh}.

     Next let us
 introduce a conjugate fermionic fields $\eta(z)$ and $\xi(z)$ by
\begin{eqnarray}
    \eta(z)&=&\exp\left\{\bosc{2}{q^{-k-2}z}{0}\right\}, \\
    \xi(z)&=&\exp\left\{-\bosc{2}{q^{-k-2}z}{0}\right\}.
\end{eqnarray}
One can show that the zero mode of the field $\eta(z)$,
$\eta_0\equiv \oint\frac{dz}{2\pi i}\eta(z)$ commutes with all the generators
in $\uqa$\cite{Ma,Koa}.

    The $q$-Wakimoto module $W_l$ is then defined from $F_l$ by the projection
onto  the kernel of the operator $\eta_0$.
\begin{equation}
   W_l=\oplus_{s\in \bf Z}{\rm Ker}\ \eta_0 (F_{l,s,s}).
\end{equation}

\vskip 2mm
\noindent{\bf 2.4 Definition of $q$-vertex operators} \quad

There are two types of the $q$-vertex operators\cite{DFJMN}.
Let $V(\lambda)$
be a highest weight $\uqa$ module of weight $\lambda$.
The type I and the type II $q$-vertex operators ($q$-VOs)
are the intertwiner

\noindent{Type I:}
\begin{eqnarray}
   \Phi_{\lambda}^{\mu V^{(l)}} (z) &=& z^{\Delta_{\mu} - \Delta_{\lambda}}
   \tilde{\Phi}_{\lambda}^{\mu V^{(l)}} (z) \nonumber \\
   \tilde{\Phi}_{\lambda}^{\mu V^{(l)}} (z) &:&
   V(\lambda) \longrightarrow V(\mu)\widehat{\otimes}V_{z}^{(l)},
   \label{VO1}
\end{eqnarray}

\noindent{Type II:}
\begin{eqnarray}
   \Phi_{\lambda}^{V^{(l)}\mu} (z) &=& z^{\Delta_{\mu} - \Delta_{\lambda}}
   \tilde{\Phi}_{\lambda}^{ V^{(l)}\mu} (z) \nonumber \\
   \tilde{\Phi}_{\lambda}^{ V^{(l)}\mu} (z) &:&
   V(\lambda) \longrightarrow V_{z}^{(l)}\widehat{\otimes}V(\mu).
   \label{intwt}
\end{eqnarray}

The $q$-VOs satisfy the intertwining relations
\begin{eqnarray}
   \tilde{\Phi}_{\lambda}^{\mu V^{(l)}} (z) \circ x &=&
   \Delta (x) \circ \tilde{\Phi}_{\lambda}^{\mu V^{(l)}} (z),
   \label{intertwine} \\
   \tilde{\Phi}_{\lambda}^{V^{(l)}\mu} (z) \circ x &=&
   \Delta (x) \circ \tilde{\Phi}_{\lambda}^{V^{(l)}\mu} (z),\qquad
   \forall x \in \uqa,
   \label{tintertwine}
\end{eqnarray}
where $\triangle(x)$ denotes a comultiplication:
\begin{eqnarray}
   \triangle (e_i)&=&e_i\otimes 1+ t_i\otimes e_i, \qquad
   \triangle (f_i)=f_i\otimes t_i^{-1}+ 1\otimes f_i, \nonumber \\
   \triangle (t_i)&=& t_i\otimes t_i.
\label{comch}
\end{eqnarray}

    According to Ref.\cite{IIJMNT}, we take
the normalization of the type I and the type II $q$-VOs
as follows\footnote{
In the most part of this paper, we consider only the ``spin $1/2$''
type II $q$-VO $\Phi^{V^{(1)}\lambda}_{\mu}(z)$. This is due
to the fact that even in the spin $k/2$ XXZ model, only the spin $1/2$
physical particle appears\cite{IIJMNT}.}.
\begin{equation}
   \tilde{\Phi}_{\lambda}^{\mu V^{(l)}} (z) \ket{\lambda} =
   \ket{\mu} \otimes v_{m}^{(l)} + \cdots,
   \label{norm}
\end{equation}
for
$\lambda=\lambda_n$ and
 $\mu=\sigma(\lambda_n)\equiv n\Lambda_0+(k-n)\Lambda_1$ ,
and
\begin{equation}
   \tilde{\Phi}_{\lambda}^{V^{(1)}\mu} (z) \ket{\lambda} =
   v_{\mp}^{(1)} \otimes \ket{\mu} + \cdots,
   \label{tnorm}
\end{equation}
for
$\lambda=\lambda_n$ and $\mu=\lambda_{\pm}\equiv \lambda_{n\pm1}$,
respectively.

     The free field realization of the $q$-VOs can be determined by the
intertwining relations (\ref{intertwine}) and (\ref{tintertwine})
with the
 in (\ref{comch})  as well as those reexpressed by
the Drinfeld's generators\cite{CP}:

$$\begin{array}{l} \left.
   \begin{array}{lcl} \Delta(J_{k}^{+}) &=&
   J_{k}^{+}\otimes\gamma^{k}+\gamma^{2k}K\otimes J_k^{+}+ \displaystyle
   \sum_{i=0}^{k-1}\gamma^{(k+3i)/2}\psi_{k-i}\otimes\gamma^{k-i}J_i^{+}
    \\
   \Delta(J_{-l}^{+})&=& J_{-l}^{+}\otimes\gamma^{-l}+K^{-1}\otimes
   J_{-l}^{+}+ \displaystyle \sum_{i=1}^{l-1}\gamma^{(l-i)/2}\varphi_{-l+i}
   \otimes\gamma^{-l+i}J_{-i}^{+}
   \end{array} \right\} \bmod N_{-}\otimes
   N_{+}^2,\hh
\end{array}
$$
\begin{equation}
\begin{array}{l}
   \left.  \begin{array}{lcl} \Delta(J_{l}^{-}) &=&
   J_{l}^{-}\otimes K + \gamma^{l}\otimes J_{l}^{-} + \displaystyle
   \sum_{i=1}^{l-1}\gamma^{l-i}J_{i}^{-}\otimes\gamma^{(i-l)/2}\psi_{l-i}
   \\
   \Delta(J_{-k}^{-}) &=& J_{-k}^{-}\otimes\gamma^{-2k} K^{-1} +
   \gamma^{-k}\otimes J_{-k}^{-}+ \displaystyle
   \sum_{i=0}^{k-1}\gamma^{i-k}J_{-i}^{-}
   \otimes\gamma^{-(k+3i)/2}\varphi_{i-k} \end{array} \right\} \bmod
   N_{-}^2\otimes N_{+}, \hh \nonumber\\
   \left.
   \begin{array}{lcl}
   \Delta(J^{3}_l)&=&J^{3}_l\otimes \gamma^{l/2}+\gamma^{3l/2}\otimes
   J^{3}_l \hh \\
   \Delta(J^{3}_{-l})&=&J^{3}_{-l}\otimes\gamma^{-3l/2}
   +\gamma^{-l/2}\otimes J^{3}_{-l} \hh \end{array} \right\} \bmod
   N_{-}\otimes N_{+}\\
 \end{array}
\label{comulti}
\end{equation}
for $k \in \bz_{\geq0}$ and $l \in \bz_{>0}$.
Here $N_{\pm}$ and $N_{\pm}^2$ are left
${\bf Q}(q)[\gamma^\pm,\psi_r,\varphi_s|r,-s\in\bz_{\geq0}]$-modules
 generated by $\{J_m^{\pm}| m \in \bz \}$ and
$\{J_m^{\pm}J_n^{\pm}|m,n\in \bz\}$ respectively.

        According to Refs.\cite{Sh,Ma,Koa}, the free field realization
is given for the
type I $q$-VO as follows.
\begin{equation}
\tilde{\Phi}_{\lambda n}^{\lambda_{n'}V^{(l)}}(z)=
g^{\lambda_{n'}V^{(l)}}_{\lambda_n}(z)
\sum_{m=0}^l\Psi_{l,m}^{(r)}(z)\otimes v^{(l)}_m,
\label{deftypeI}
\end{equation}
\begin{equation}
\Psi^{(r)}_{l,m}(z)=
              \jint{q^{-l}z}{1}\jint{q^{-l}z}{2}\cdots \jint{q^{-l}z}{r}
             \Psi_{l,m}(z)S(t_1)S(t_2)\cdots S(t_r),
               \label{defsvo}
\end{equation}
\begin{eqnarray}
\Psi_{l,m} (z)&=&
               \frac{[m]}{[l]!}\oint dw_{1}\oint dw_{2}\cdots\oint dw_{l-m}
          \nonumber \\
               && \qquad \qquad \times
    [\cdots[~[ \Psi_{l,l}(z), J^{-}(w_{1}) ]_{q^l}, J^{-}(w_{2}) ]_{q^{l-2}}
               \cdots  J^{-}(w_{l-m}) ]_{q^{2(m+1)-l}},
\end{eqnarray}
\begin{equation}
\Psi_{l,l}(z)= ~~: \exp \Bigl\{
        \bosal{l}{2}{k+2}{q^{k}z} {\tfra{k+2}{2}} \Bigr\} :,
   \label{eVOdef1}
\end{equation}
where $2r=n+l-n'$, and  $g^{\lambda_{n'}V^{(l)}}_{\lambda_n}(z)$
being the normalization
function.
In (\ref{defsvo}), we introduced
the screening operator
 ${S}(z)$, which commutes with $\uqa$ and $\eta_0$ up to a total
difference term. It is explicitly given by
\begin{equation}
\begin{array}{rl}
S(z) =& - :\left[ \diff{1}{z}
\exp \left\{ -\bosc{2}{q^{-k-2}z}{0} \right\} \right] \hh \\
& \times \exp \left\{ -\bosb{2}{q^{-k-2}z}{-1}
- \bosa{k+2}{q^{-2}z}{-\frac{k+2}{2}} \right\} : . \hh
\end{array}
\end{equation}
Here, the symbol  $\jint{c}{}$ denotes
the Jackson integral  defined by
\begin{equation}
\jint{c}{} f(t) = c(1-p) \displaystyle
\sum^{\infty}_{m=0} f(cp^{m})p^{m},
\label{jackson}
\end{equation}
for  $p \in \bc$, $|p|<1$,
and $c\in \bc^{\times}$.
In our case $p\equiv q^{2(k+2)}$.
   The free field realization of the type II $q$-VO is discussed  in the
next section.

\vskip 2mm
\noindent{\bf 2.5 BRST-Felder cohomology } \par
      Let   $n\geq 1 ,n'\geq 0$ be integers,
 and $P,P'$ be  coprime positive integers satisfying $\frac{P}{ P'}= k+2$.
Then the $q$-Wakimoto module $W_{n,n'}\equiv W_{l_{n,n'}}$ with
$l_{n,n'}=n-n'\frac{P}{ P'}-1$ is reducible\cite{Koa}.
We are especially interested in the case $1\leq n\leq P-1$ and
$0\leq n'\leq P'-1$.
For
this case,  the resolution of IHWR can be given as a cohomology group
in a BRST like complex of the $q$-Wakimoto modules\cite{Koa}.

     Let us define the BRST charge $Q_n$, $n\in {\bf Z}_{>0}$, by
\begin{eqnarray}
Q_n&=&\int_{0}^{c\infty}d_pt\int_{0}^{q^2t\infty}d_pt_2\cdots
\int_{0}^{q^2t\infty}d_pt_n
         S(t)S(t_2)\cdots S(t_n),
\end{eqnarray}
where $c\in \bc^{\times}$ is arbitrary.

     The BRST charge $Q_n$ satisfies the following properties.

\noindent{(i)\quad $ Q_n$ commutes with $\uqa$ and $\eta_0$.}

\noindent{(ii)\ $ Q_nQ_{P-n}=Q_{P-n}Q_n=0.$}

\noindent{(iii)The following infinite sequence}
\begin{equation}
\cdots
{\buildrel Q_{n}\over\longrightarrow}W_{-n+2P,n'}
{\buildrel Q_{P-n}\over\longrightarrow}W_{n,n'}
{\buildrel Q_{n}\over\longrightarrow}W_{-n,n'}
{\buildrel Q_{P-n}\over\longrightarrow}W_{n-2P,n'}
{\buildrel Q_{n}\over\longrightarrow}W_{-n-2P,n'}
{\buildrel Q_{P-n}\over\longrightarrow}
\cdots
\label{complex}
\end{equation}
is a complex. \par
\par
     Constructing the singular and
cosingular vectors in the $q$-Wakimoto module $W_{n,n'}$, one can show
 that the
$\uqa$ submodule
structure generated by the singular and cosingular
vectors in the complex has the same structure
as the classical case ($q$=1)\cite{Koa}( see also \cite{Mali,FeFr}).

     This structure gives rise to the following
cohomology groups of  the complex (\ref{complex})
\begin{equation}
{\rm Ker} Q_n^{[s]}/{\rm Im} Q_n^{[s-1]}
=\Bigl\lbrace\matrix{0&for&s\not=0\cr
               {\cal H}_{n,n'}&for&s=0\cr},
\end{equation}
where $Q^{[2a]}_n=Q_n$ and $Q_n^{[2a-1]}=Q_{P-n}$ with $a\in \bf Z$
and the space ${\cal H}_{n,n'}$ is the IHWR for  $\uqa$
of highest weight $\lambda_{l_{n,n'}}$.
\par
     As a by-product, we obtain a formula for the trace over the IHWR of
$\uqa$
\begin{equation}
\Tr_{{\cal H}_{n,n'}}{\cal O}=\sum_{s\in \bf Z}(-)^s
\Tr_{{W}^{[s]}_{n,n'}}{\cal O}^{[s]},
\label{trihwr}
\end{equation}
where the graded physical operator ${\cal O}^{[s]}$ is defined recursively by
the relations

\begin{equation}
Q_n^{[s]}{\cal O}^{[s]}={\cal O}^{[s+1]}Q_n^{[s]},\quad
{\cal O}^{[0]}={\cal O}.
\end{equation}

Furthermore, noting the defining
process of the
$q$-Wakimoto modules from the Fock modules,
we relate  the traces over the $q$-Wakimoto modules to those on
the Fock modules:
\begin{equation}
\Tr_{W^{[s]}_{n,n'}}{\cal O}^{[s]}
=\Tr_{F^{[s]}_{n,n'}}\Bigl(\oint{dw_0\over 2\pi i w_0}\xi(w_0)
\oint{dw\over 2\pi i}\eta(w)
{\cal O}^{[s]}\Bigr)
\Bigl|_{\tilde a_{\phi,0}+\tilde a_{\chi,0}=0},
\label{trwakimoto}
\end{equation}
where the restriction $\tilde a_{\phi,0}+\tilde a_{\chi,0}=0$ should be
understood on its eigenvalues.
The RHS can be evaluated in terms of the free fields.

\section{Type II $q$-Vertex Operator}
In this section,  we discuss a free field realization
of the type II $q$-vertex
operator.

\vskip 2mm
\noindent{\bf 3.1 Free field realization } \quad

In the same way as the type I case\cite{Sh},
the key to realizing the type II
$q$-VO  is
the intertwining relation (\ref{tintertwine}) with the
comultiplication formulae (\ref{comch}) and (\ref{comulti}).

     Let us write the type II $q$-VO as follows.
\begin{equation}
\tilde \Phi^{V^{(l)}\mu}_{\lambda}(z)=\sum_{m=0}^l v^{(l)}_m\otimes
                          \tilde \Phi^{V^{(l)}\mu}_{\lambda,m}(z).
\label{decompo}
\end{equation}
Then, from (\ref{tintertwine}) and (\ref{comch}),  we have
\begin{equation}
\tilde \Phi^{V^{(l)}\mu}_{\lambda,m+1}(z)=\fra{1}{[m+1]}\left\{
    \tilde \Phi^{V^{(l)}\mu}_{\lambda,m}(z)e_1-
    q^{l-2m}e_1 \tilde \Phi^{V^{(l)}\mu}_{\lambda,m}(z)\right\},
\label{ttlm}
\end{equation}
for $m=0,1,2,..,l-1$.

    Furthermore,
by using (\ref{dfinitem}), (\ref{tintertwine}) and (\ref{comulti}),
we obtain the intertwining relations for
$\tilde \Phi^{V^{(l)}\mu}_{\lambda,0}(z)$ as follows.
\begin{eqnarray}
   \hbox{$[J^{3}_n,\tilde{\Phi}_{\lambda ,0}^{V^{(l)}\mu}(z)]$}&=&
   - z^{n} \frac{[nl]}{n}
   \cdot q^{k(n-|n|/2)}
   \tilde{\Phi}_{\lambda ,0}^{V^{(l)}\mu}(z) \qquad n \neq 0,
   \nonumber
\\
   \hbox{$[\tilde{\Phi}_{\lambda ,0}^{V^{(l)}\mu}(z),J^{-}(w)]$}&=&0,
   \label{commt}
\\
   K \tilde{\Phi}_{\lambda,0}^{V^{(l)}\mu}(z) K^{-1}
    &=& q^{-l}\tilde{\Phi}_{\lambda,0}^{V^{(l)}\mu}(z).
   \nonumber
\end{eqnarray}
These follow from $ v^{(l)}_{0} \otimes V(\mu)$ components
of the intertwining relation.

We find that the following vertex operator satisfies the whole relations
in (\ref{commt})
\begin{eqnarray}
   \Phi_{l,0}(z)&=& : \exp \Bigl\{
    \bosal{l}{2}{k+2}{q^{k-2}z}{-\tfra{k+2}{2}} \nonumber \\
&&\qquad +
    \bosbl{l}{2}{1}{q^{-2}z}{0}+
    \boscl{l}{2}{1}{q^{-2}z}{0} \Bigr\} :.
   \label{tVOdef1}
\end{eqnarray}

 From (\ref{ttlm}), we then obtain
\begin{eqnarray}
\lefteqn{
   \Phi_{l,m} (z) =
   \frac{1}{[m]!}
   \oint \fra{du_{1}}{2\pi i}
   \oint \fra{du_{2}}{2\pi i}
   \cdots
   \oint \fra{du_{m}}{2\pi i}
   }
   \nonumber
   \\
   && \qquad \qquad \times
   [\cdots[~[ \Phi_{l,0}(z), J^{+}(u_{1}) ]_{q^l}, J^{+}(u_{2}) ]_{q^{l-2}}
    \cdots  J^{+}(u_{m}) ]_{q^{l-2(m-1)}} ,
   \label{ttlmop}
\end{eqnarray}
for $m=1,2,...,l$.

Noting $[\Phi_{l,m}(z), \tilde\aphi{0}+\tilde\achi{0}]=0$
and $[\Phi_{l,m}(z), \eta_{0}]=0$, the vertex operator
$\Phi_{l,m}(z)$ implies a linear map $\Phi_{l,m}(z): W_{l'}\to W_{l'+l}$.
Further, noting  the fact $S(t): W_{l}\to W_{l-2}$, we
obtain a linear map
\begin{equation}
  \jintinf{1}\jintinf{2}\cdots\jintinf{r}\Phi_{l,m}(z)S(t_1)S(t_2)\cdots
              S(t_r): W_{l'}\to W_{l''},
\end{equation}
with $2r=l'+l-l''$.
Since the screening charge $\jintinf{}S(t)$ commutes with
the operator $\tilde\aphi{0}+\tilde\achi{0}, \quad \eta_0$ and
all the generators
in $\uqa$,
we obtain the following identification:
\begin{eqnarray}
    {\tilde\Phi}^{V^{(l)}\lambda_{l''}}_{\lambda_{l'},m}(z)&=&
                  g^{V^{(l)}\lambda_{l''}}_{\lambda_{l'}}(z)
                   {\Phi}^{(r)}_{l,m}(z), \label{ttqvo}\\
    {\Phi}^{(r)}_{l,m}(z)
       & =&
   \int_{0}^{c\infty} d_{p} t_{1}
   \int_{0}^{c\infty} d_{p} t_{2}
   \cdots
   \int_{0}^{c\infty} d_{p} t_{r}
   \Phi_{l,m}(z){S}(t_{1})S(t_2)\cdots  {S}(t_{r}).
\label{ttscv}
\end{eqnarray}
Here $g^{ V^{(l)}\lambda_{l''}}_{\lambda_l'}(z)$
is the normalization factor to be determined in the below.

\vskip 2mm
\noindent{\bf 3.2 Cycles for the Jackson integral } \quad

    The expression (\ref{ttscv}) is not well defined in the regions
$|t_i/z|>1, i=1,2,..,r$.
We here determine suitable cycles, which make sense to  the Jackson integrals
in (\ref{ttscv}).  Our argument is simply based on the requirement of
the commutativity of the screening charge
with the currents in $\uqa$.   Remarkably, it turns out that the resultant
cycles imply  the
$R$-matrix symmetry to  the type II $q$-VOs.
As will be shown in Sec.5, this symmetry
is essential in the derivation of the properties of the form factors.

     We have to consider the two cases $\int d_pt\ \Phi_{l,0}(z)S(z)$ and
$\int d_pt\ S(t)\Phi_{l,0}(z)$.  As discussed in Ref\cite{Koa},
the only  harmful operator
in the commutativity with $S(t)$ is the current $J^-(w)$, whose operator
expansion with $S(t)$ is given by
\begin{eqnarray}
S(t)J^-(w)&=&J^-(w)S(t)\nonumber \\
          &\sim&\diff{k+2}{t}
          \Bigl[\frac{1}{w-t}:e^{-\bosa{k+2}{q^{-2}t}{\frac{k+2}{2}}}:\Bigr].
\end{eqnarray}
Therefore, in the commutation of
$\int d_pt\Phi_{l,0}(z)S(t)$ with $J^-(w)$,  we have a Jackson integral
of the total difference $\diff{k+2}{t}$
of the function containing the factor
\begin{equation}
\Phi_{l,0}(z):e^{-\bosa{k+2}{q^{-2}t}{\frac{k+2}{2}}}:\
\sim \ \frac{\inp{q^{l+2}t/z}{p}}{\inp{q^{-l+2}t/z}{p}}
         : \Phi_{l,0}(z)e^{-\bosa{k+2}{q^{-2}t}{\frac{k+2}{2}}}:.
\label{totdiffa}
\end{equation}
In the same way, for $\int d_pt S(t)\Phi_{l,0}(z)$, we have the  total
difference of
the function containing
\begin{equation}
:e^{-\bosa{k+2}{q^{-2}t}{\frac{k+2}{2}}}:\Phi_{l,0}(z)\
\sim \ \frac{\inp{q^{l+2k+2}z/t}{p}}{\inp{q^{-l+2k+2}z/t}{p}}
         : \Phi_{l,0}(z)e^{-\bosa{k+2}{q^{-2}t}{\frac{k+2}{2}}}:.
\label{totdiffb}
\end{equation}

      Now let us look for the finite regions,
by which the Jackson integrals of
the total differences vanish. By definition of the Jackson integral
(\ref{jackson}),
it is  easy to find that
the Jackson integral
 of the total difference of (\ref{totdiffa}) over any one of the regions
$[0, q^{k-l}p^{-j}z],\quad j=0,1,2..$  vanishes.
Similarly,  we find that any one of the regions
$[q^{k+l}p^{j+1}z,q^{k+l}p^j z\infty),\quad j=0,1,2,..$ makes the total
difference of (\ref{totdiffb}) vanish.

     Let us consider in detail
the physically interesting case $l=1$:
\begin{equation}
\tilde\Phi^{V^{(1)}\mu}_{\lambda}(z)=
v_+\otimes\tilde\Phi^{V^{(1)}\mu}_{\lambda,0}(z)
+v_-\otimes\tilde\Phi^{V^{(1)}\mu}_{\lambda,1}(z),
\end{equation}
with $\mu=\lambda_{\pm}$.

    For $\mu=\lambda_+$, we have $r=0$ in (\ref{ttqvo}),
and obtain the expression
\begin{equation}
\tilde\Phi^{V^{(1)}\lambda_+}_{\lambda,0}(z)=
g^{V^{(1)}\lambda_+}_{\lambda}(z)\Phi_{1,0}(z).
\end{equation}
For $\mu=\lambda_-$, we have $r=1$ in (\ref{ttqvo}) and obtain,
from the above arguments, the following
two expressions
for the vertex  $\tilde\Phi^{V^{(1)}\lambda_-}_{\lambda,0}(z)$:
\begin{equation}
g^{ V^{(1)}\lambda_-}_{\lambda}(z)
\jint{q^{k-1}z}{}\Phi_{1,0}(z)S(t), \\
\label{svodn}
\end{equation}
\begin{equation}
\tilde g^{ V^{(1)}\lambda_-}_{\lambda}(z)
\jintd{q^{k+1}pz}{q^{k+1}z}{}S(t)\Phi_{1,0}(z),
\label{svoup}
\end{equation}
in the simplest choice $j=0$.

   The normalization functions can be easily calculated using
the expression (\ref{tVOdef1}) and operator product
expansion formulae in appendix A.
 From (\ref{tnorm}) and (\ref{decompo}), we obtain the following results
for the VOs $\tilde \Phi^{V^{(1)}\lambda_{l+1}}_{\lambda_{l}}(z)$ and
$\tilde \Phi^{V^{(1)}\lambda_{l-1}}_{\lambda_{l}}(z)$:
\begin{eqnarray}
g^{ V^{(1)}\lambda_{l+1}}_{\lambda_l}(z)&=&q^{-1-2l(k-2)s}z^{-2ls},\\
g^{ V^{(1)}\lambda_{l-1}}_{\lambda_l}(z)&=&-q^{2+l/2- 2(3l+4)s}z^{(l+2)s}
               B_p(1-2ls,-2s)^{-1}.
\end{eqnarray}
Here $s\equiv \frac{1}{2(k+2)}$, and
$$B_p(x,y)=\int_0^1 d_pt \ t^{x-1}\frac{(pt;p)_{\infty}}{(p^yt;p)_{\infty}}$$
being the $q$-beta function\cite{GR}.
It is useful to notice that in terms of the $q$-gamma function defined by
$$\pGam{z}= \frac{\inp{p}{p}}{\inp{p^z}{p}}(1-p)^{1-z},$$
the $q$-beta function is expressed by
$$B_p(x,y)=\frac{\Gamma_p(x)\Gamma_p(y)}{\Gamma_p(x+y)}.$$

In the same way,  from (\ref{tnorm}),
(\ref{decompo}) and (\ref{svoup}),  we obtain
\begin{equation}
\tilde g^{ V^{(1)}\lambda_{l-1}}_{\lambda_l}(z)=-q^{-1+l/2- 2(l+3)s}z^{(l+2)s}
               B_p(2(l+1)s,-2s)^{-1}.
\end{equation}

\vskip 2mm
\noindent{\bf 3.3 Commutation relation } \quad

    We next discuss a commutation relation of the type II $q$-VOs. We show that
the cycles obtained in the above lead to the $R$-matrix symmetry.\par
    For this purpose, we calculate
the following two point functions.
\begin{equation}
<\lambda|\tilde \Phi^{V^{(1)}_2 \lambda}_{\mu}(z_2)
\tilde \Phi^{V^{(1)}_1 \mu}_{\lambda}(z_1)|\lambda>
=
\sum_{m_1,m_2=0,1 \atop m_1+m_2=1}v^{(1)}_{m_1}\otimes v^{(1)}_{m_2}
<\lambda|\tilde\Phi^{V^{(1)}_2\lambda}_{\mu,m_2}(z_2)
\tilde \Phi^{V^{(1)}_1 \mu}_{\lambda,m_1}(z_1)|\lambda> .
\label{tfnc}
\end{equation}

    Note that
in (\ref{tfnc}) only the cases $\mu=\lambda_{\pm}$ give non-zero contributions.
Let us first consider the case $\mu=\lambda_-$.

     For the sake of brevity, let us  set $\Phi^{\mu}_{\lambda,+}(z)=
\tilde \Phi^{V^{(1)} \mu}_{\lambda,0}(z)$ and  $\Phi^{\mu}_{\lambda,-}(z)=
\tilde \Phi^{V^{(1)} \mu}_{\lambda,1}(z)$.
By the direct
calculation using (\ref{svodn}) for $\Phi^{\lambda_-}_{\lambda,+}(z)$
and the operator expansion formulae
given in Appendix A, we obtain  the following results, for $\lambda=\lambda_l$.
\begin{eqnarray}
\lefteqn{
<\lambda|\Phi^{\lambda}_{\lambda_-,+}(z_2)
\Phi^{ \lambda_-}_{\lambda,-}(z_1)|\lambda>/\xi(z;1,q^4)
}
\nonumber \\
&=& g^{V^{(1)}\lambda}_{\lambda_-}(z_2)
g^{V^{(1)}\lambda_-}_{\lambda}(z_1)
\jint{q^{k-1}z_1}{} \oint \fra{du}{2\pi i}
<\lambda|\Phi_{1,0}(z_2)[\Phi_{1,0 }(z_1),J^+(u)]_q S(t)|\lambda>
\nonumber \\
&=&B_p(1-2ls,-2s)^{-1}qz
\intb{-2ls}{p}{q^{-2}p},
\label{tfmmp}
\end{eqnarray}

\begin{eqnarray}
\lefteqn{
<\lambda|\Phi^{\lambda}_{\lambda_-,-}(z_2)
 \Phi^{ \lambda_-}_{\lambda,+}(z_1)|\lambda>/\xi(z;1,q^4)
}\nonumber \\
&=&g^{V^{(1)}\lambda}_{\lambda_-}(z_2)
g^{V^{(1)}\lambda_-}_{\lambda}(z_1)
\jint{q^{k-1}z_1}{}\oint \fra{du}{2\pi i}
<\lambda|[\Phi_{1,0}(z_2),J^+(u)]_q\Phi_{1,0 }(z_1) S(t)|\lambda>
\nonumber \\
&=&B_p(1-2ls,-2s)^{-1}
\intb{-2ls}{}{q^{-2}},
\label{tfmpm}
\end{eqnarray}
where  $z=z_1/z_2$.

      Here the function $\xi(z;1,q^4)$ is given by
\begin{equation}
\xi(z;a,b)={\indp{az}{p}{q^4}\indp{a^{-1}bz}{p}{q^4}\over
\indp{q^2az}{p}{q^4}\indp{q^{-2}a^{-1}bz}{p}{q^4}}.
\end{equation}

Combining (\ref{tfmpm}) and (\ref{tfmmp}), we obtain
\begin{eqnarray}
\lefteqn{
<\lambda|\tilde \Phi^{V^{(1)}_2 \lambda}_{\lambda_{-}}(z_2)
\tilde \Phi^{V^{(1)}_1 \lambda_{-}}_{\lambda}(z_1)|\lambda>
}\nonumber \\
&& =
\Fp{1-2ls,2s,1-2(l+1)s;q^{-2}z} v_+\otimes v_-        \nonumber \\
&& +
\fra{ \pGam{1-2(l+1)s} \pGam{1-2s} }{ \pGam{2-2(l+1)s}\pGam{-2s} }qz~
\Fp{1-2ls,1+2s,2-2(l+1)s;q^{-2}z} v_-\otimes v_+.
\label{tfm}
\end{eqnarray}
Here we used the basic hypergeometric series
$\Fp{a,b,c,:z}$ having
the following Jackson integral presentation\cite{GR}.
\begin{equation}
\Fp{a,b,c;z}=\fra{\pGam{c}}{\pGam{a}\pGam{c-a}}\defintb{a-1}{b}{c-a}.
\end{equation}

       Next let us consider the case $\mu=\lambda_+$.
In this case, using (\ref{svoup})
for $
\tilde \Phi^{V^{(1)} \lambda}_{\lambda_+,+}(z)$,
 we  obtain
\begin{eqnarray}
\lefteqn{
<\lambda|\Phi^{\lambda}_{\lambda_+,+}(z_2)
\Phi^{\lambda_+}_{\lambda,-}(z_1)|\lambda>
/\xi(z;1,q^4)
}\nonumber \\
&=&\tilde g^{V^{(1)}\lambda}_{\lambda_+}(z_2)
\tilde g^{V^{(1)}\lambda_+}_{\lambda}(z_1)
\jintd{q^{k+1}pz_2}{q^{k+1}z_2}{}\oint \fra{du}{2\pi i}
<\lambda|S(t)\Phi_{1,0}(z_2)[\Phi_{1,0 }(z_1),J^{+}(u)]_q |\lambda> \nonumber\\
&=&B_p(2(l+2)s,-2s)^{-1}q^{-2}
\intbd{-2(l+2)s-1}{}{q^{-2}}\nonumber \\
&=&B_p(2(l+2)s,-2s)^{-1}q^{-2}
\intb{2(l+2)s-1}{}{q^{-2}}.
\label{tfppm}
\end{eqnarray}
In the last line, we made a change $t\rightarrow {1}/{t}$ by the formula
\begin{equation}
\int_p^{\infty}d_pt\ f(t)=\jint{1}{}t^{-2}f(1/t).
\end{equation}

        Similarly, we obtain
\begin{eqnarray}
\lefteqn{
 <\lambda|\Phi^{\lambda}_{\lambda_+,-}(z_2)
 \Phi^{ \lambda_+}_{\lambda,+}(z_1)|\lambda>/\xi(z;1,q^4)
    } \nonumber \\
&&=\tilde g^{V^{(1)}\lambda}_{\lambda_+}(z_2)
\tilde g^{V^{(1)}\lambda_+}_{\lambda}(z_1)
\jintd{q^{k+1}pz_2}{q^{k+1}z_2}{}\oint \fra{du}{2\pi i}
<\lambda|S(t)[\Phi_{1,0}(z_2),J^+(u)]_q\Phi_{1,0 }(z_1)|\lambda>
       \nonumber \\
&&=B_p(2(l+2)s,-2s)^{-1}q^{-1}
\intb{2(l+2)s-1}{p}{q^{-2}p}.
\label{tfpmp}
\end{eqnarray}

Combining (\ref{tfppm}) and (\ref{tfpmp}), we find
\begin{eqnarray}
\lefteqn{
<\lambda|\tilde \Phi^{V^{(1)}_2 \lambda}_{\lambda_{+}}(z_2)
\tilde \Phi^{V^{(1)}_1 \lambda_{+}}_{\lambda}(z_1)|\lambda>
}\nonumber \\
&&=
F_p(2(l+2)s,2s,2(l+1)s;q^{-2}z)v_-\otimes v_+
 \nonumber \\
&&
+\frac{\pGam{2(l+1)s}\pGam{1-2s}}{\pGam{1+2(l+1)s}\pGam{-2s}}q
\Fp{2(l+2)s,1+2s,1+2(l+1)s;q^{-2}z}v_+\otimes v_-.
\label{tfp}
\end{eqnarray}

    The results (\ref{tfm}) and (\ref{tfp}) coincide  with those obtained in
Ref.\cite{IIJMNT}, where no
specific realizations of the $q$-VO's are used.

     Using these results as well as the connection formula for
the basic hypergeometric series\cite{Mim} and the $q$-KZ
equation for the two point function, one  obtains
the commutation relation of the type II $q$-VO
(the $R$-matrix symmetry)\cite{IIJMNT}:
\begin{equation}
\Phi^{\nu}_{\lambda_{\pm},\varepsilon_1}(z_1)
\Phi^{\lambda_{\pm}}_{\lambda,\varepsilon_2}(z_2)
=R_{\varepsilon_1\varepsilon_2}^{\varepsilon'_1\varepsilon'_2}(z)
\sum_{\mu=\lambda_+,\lambda-}
             \Phi^{\lambda}_{\mu,\varepsilon'_2}(z_2)
            \Phi^{ \mu}_{\lambda,\varepsilon'_1}(z_1)
            W\left(\begin{array}{cc}
                    \lambda&\lambda_{\pm}\\
                            \mu&\nu
                    \end{array}{\Biggl |} z\right),
\label{crttvo}
\end{equation}
where $\varepsilon_i=\pm, i=1,2$, and
the coefficients
$R_{\varepsilon_1\varepsilon_2}^{\varepsilon'_1\varepsilon'_2}(z)$
are the  $R$-matrix, $R(z):V^{(1)}\otimes V^{(1)}
\rightarrow V^{(1)}\otimes V^{(1)}$
given by
\begin{eqnarray}
R_{\varepsilon_1\varepsilon_2}^{\varepsilon'_1\varepsilon'_2}(z)
&=&r(z)r_{\varepsilon_1\varepsilon_2}^{\varepsilon'_1\varepsilon'_2}(z),\qquad
r(z)=z^{-1/2}\fra{\inp{q^4/z}{q^4}\inp{q^2 z}{q^4}}
{\inp{q^4z}{q^4}\inp{q^2/ z}{q^4}},\\
r_{++}^{++}(z)&=&r_{--}^{--}(z)=1,\\
r_{+-}^{+-}(z)&=&r_{-+}^{-+}(z)=\fra{1-z}{1-q^2z}q,\\
r_{+-}^{-+}(z)&=&zr_{-+}^{+-}(z)=\fra{1-q^2}{1-q^2z}z.
\label{Rmatrix}
\end{eqnarray}
The factors $W\left(\matrix{\lambda&\mu\cr
                            \mu'&\nu \cr}{\Biggl |}z\right)$ are
given by
\begin{eqnarray}
W\left(\matrix{\lambda&\mu\cr
                            \mu'&\nu\cr}{\Biggl |}z\right)
&=&-z^{\triangle_{\lambda}+\triangle_{\nu}-\triangle_{\mu}-\triangle_{\mu'}-1/2}
\frac{\xi(z^{-1};1,pq^4)}{\xi(z;1,pq^4)}
{\hat W}\left(\matrix{\lambda&\mu\cr
                            \mu'&\nu\cr}{\Biggl |}z\right),\\
{\hat W}\left(\matrix{\lambda&\lambda_{+}\cr
                            \lambda_+&\lambda \cr}{\Biggl |}z\right)&=&
\frac{\Theta_p(pq^2)}{\Theta_p(pq^{-2l-2})}
\frac{\Theta_p(pq^{-2l-2}z)}{\Theta_p(pq^{2}z)},\\
{\hat W}\left(\matrix{\lambda&\lambda_{+}\cr
                            \lambda_-&\lambda \cr}{\Biggl |}z\right)&=&
q^{-1}\frac{\Gamma_p((2l+2)s)^2}{\Gamma_p((2l+4)s)\Gamma_p(2ls)}
\frac{\Theta_p(pz)}{\Theta_p(pq^{2}z)},\\
{\hat W}\left(\matrix{\lambda&\lambda_{-}\cr
                            \lambda_+&\lambda \cr}{\Biggl |}z\right)&=&
q^{-1}\frac{\Gamma_p(1-(2l+2)s)^2}{\Gamma_p(1-(2l+4)s)\Gamma_p(1-2ls)}
\frac{\Theta_p(pz)}{\Theta_p(pq^{2}z)},\\
{\hat W}\left(\matrix{\lambda&\lambda_{-}\cr
                            \lambda_-&\lambda \cr}{\Biggl |}z\right)&=&
z^{-1}\frac{\Theta_p(pq^2)}{\Theta_p(q^{2l+2})}
\frac{\Theta_p(q^{2l+2}z)}{\Theta_p(pq^{2}z)},\\
{\hat W}\left(\matrix{\lambda&\lambda_{\pm}\cr
               \lambda_{\pm}&\lambda_{\pm}^{\pm} \cr}{\Biggl |}z\right)&=& 1,\\
{\hat W}\left(\matrix{\lambda&\mu\cr
                            \mu'&\nu \cr}{\Biggl |}z\right)&=& 0 \qquad
{\rm otherwise}.
\label{wweight}
\end{eqnarray}
Here $\Theta_p(z)=\inp{z}{p}\inp{z^{-1}p}{p}\inp{p}{p}$.
These factors can be regarded as the Boltzmann weight
in the face formulation.

\section{ Integral Formula for Form Factors}

In this section,
we
evaluate a trace of the $q$-VOs over IHWR of $\uqa$
 and derive a general  integral formula for form factors in
the spin $k/2$ antiferomagnetic XXZ model. We here assume $-1<q<0$\cite{DFJMN}.

    Let ${\cal L}_{\lambda}$ be a local operator
 acting on the $q$-Wakimoto module $W_{\lambda}$.
As discussed in  Ref.\cite{DFJMN,IIJMNT}, the operator ${\cal L}_{\lambda}$ is
given  by the product of the type I $q$-vertex operators. It enjoys the
property
\begin{equation}
\cLl\Phi^{V^{(1)}\lambda}_{\mu}(z)=\Phi^{V^{(1)}\lambda}_{\mu}(z)\cLm.
\end{equation}
In the following, we do not specify any form for  $\cLl$.

    The $N$ particle form factor of the local operator
${\cal L}_{\lambda}$ is then given by the following trace\cite{DFJMN,IIJMNT}.

\begin{eqnarray}
\lefteqn{
F_{\lambda\lambda_{N-1}\lambda_{N-2}...\lambda_1\lambda}
(\zeta_{N},\zeta_{N-1},...,\zeta_1)
}\nonumber \\
&&\qquad=\frac{\Tr_{{\cal H}_{\lambda}}\Bigl(
q^{4\tilde L_0}q^{-J^3_0}\cLl\Phi^{V^*\lambda}_{\lambda_{N-1}}(\zeta_{N})
\Phi^{V^*\lambda_{N-1}}_{\lambda_{N-2}}(\zeta_{N-1})
\cdots \Phi^{V^*\lambda_1}_{\lambda}(\zeta_1)
\Bigr)}
{\Tr_{{\cal H}_{\lambda}}\Bigl(
q^{4\tilde L_0}q^{-J^3_0}\Bigr)},
\label{Npff}
\end{eqnarray}
where $\tilde L_0\equiv L_0-{c\over 24}$ and  $c\equiv {3k\over k+2}$.
The vertex $\Phi^{V^*\lambda}_{\mu}(\zeta)
$ is an anti-pode dual of the type II $q$-vertex operator
$\Phi^{V^{(1)}\lambda}_{\mu}(z)$, describing the creation of
physical particles with spin 1/2\cite{IIJMNT}. It is related to the
type II $q$-VO as follows.

     Let us expand the vertex $\Phi^{V^*\lambda}_{\mu}(\zeta)$ as
\begin{equation}
\Phi^{V^*\lambda}_{\mu}(\zeta)=v_+\otimes \Phi^{*\lambda}_{\mu,+}(\zeta)+
v_-\otimes \Phi^{*\lambda}_{\mu,-}(\zeta),
\end{equation}
then the following relations are held\cite{IIJMNT}.
\begin{eqnarray}
 \Phi^{*\lambda_+}_{\lambda,-}(\zeta)&=&
-q\Phi^{\lambda_+}_{\lambda,+}(q^2\zeta), \label{antipda}\\
 \Phi^{*\lambda_+}_{\lambda,+}(\zeta)&=&
 \Phi^{\lambda_+}_{\lambda,-}(q^2\zeta), \\
 \Phi^{*\lambda_-}_{\lambda,-}(\zeta)&=&
 \Phi^{\lambda_-}_{\lambda,+}(q^2\zeta), \\
 \Phi^{*\lambda_-}_{\lambda,+}(\zeta)&=&
-q^{-1}\Phi^{\lambda_-}_{\lambda,-}(q^2\zeta).
\label{antipd}
\end{eqnarray}

    Note that using the $R$-matrix symmetry (\ref{crttvo})
and the cyclic property of trace, one can easily show that the form factor
(\ref{Npff}) satisfy the $q$-KZ equation.

    Now, let  us  evaluate the traces.
The calculation can be done by using the coherent state method, which is
familiar in superstring theory. For example, see Ref.\cite{AALO}. \par
     Let us first consider the normalization factor
$\Tr_{{\cal H}_{\lambda}}\Bigl(
q^{4\tilde L_0}q^{-J^3_0}\Bigr)$.
This is essentially the character of ${\cal H}_{\lambda}$.
As evaluated in Ref.\cite{Koa}, it coincides with the classical one, i.e., the
Weyl-Kac formula for $\slth$.
\begin{eqnarray}
\lefteqn{\Tr_{{\cal H}_{\lambda}}\Bigl(
q^{4\tilde L_0}q^{-J^3_0}\Bigr)}\nonumber \\
&&={1\over \vartheta_1(\tilde q^2|\tau)}
\sum_{s\in \bf Z}\Bigl[
q^{4PP'(s-{nP'-n'P\over 2PP'})^2}\tilde q^{2P(s-{nP'-n'P\over 2PP'})}
-q^{4PP'(s+{nP'+n'P\over 2PP'})^2}\tilde q^{2P(s+{nP'+n'P\over 2PP'})}\Bigr],
\end{eqnarray}
for $\lambda=\lambda_{l_{n,n'}}$,
where $\tau\equiv{\log \zeta \over 2\pi i}$,
 and $\tilde q\equiv q^{-{2\log q\over q-q^{-1}}}$.

    Next, let us consider the numerator.
 From the trace formula (\ref{trihwr}) and
(\ref{antipda}) $\sim$ (\ref{antipd}),
it is  given by a combination of the
following traces
\begin{eqnarray}
\lefteqn{
\Tr_{W_{\lambda_l}}\biggl(\zeta^{\tilde L_0}y^{-J^3_0}\prod_{i=1}^aJ^-(w_i)
\prod_{r=1}^m\Psi_{g,g}(\zeta_r)\prod_{s=1}^n\Pp{z_s}\prod_{j=1}^bS(t_j)
\prod_{h=1}^cJ^+(u_h)\biggr)
}\nonumber \\
&&=\sum_{\{\epsilon\},\{\delta\},\{\rho\}}
\fra{(-)^{b+c}\prod\epsilon_i\prod\delta_j\prod\rho_h}
{(q-q^{-1})^{a+b+c}\prod w_i\prod t_j\prod u_h}\
f_{\Phi}^{\lambda_l}(\{\epsilon\})\
f_{\phi\chi}(\{\epsilon\},\{\delta\},\{\rho\}).
\end{eqnarray}
Here we split  the operators $J^{\pm}(w) $ and $S(t)$ as follows.
\begin{eqnarray}
  J^{-}(z)&=&\fra{1}{(q-q^{-1})z}\sum_{\epsilon=\pm 1}~
                  \epsilon J^{-}_{\epsilon}(z), \\
      J^{+}(z)&=&-\fra{1}{(q-q^{-1})z}\sum_{\rho=\pm 1}~
          \rho J^{+}_{\rho}(z),\\
%
    S(z)&=&-\fra{1}{(q-q^{-1})z}\sum_{\delta=\pm 1}\delta S_{\delta}(z),
\end{eqnarray}
where
\begin{eqnarray}
   J^{-}_{\epsilon}(z)&=& :\exp \Bigl\{
          \pbosae{q^{-2}z}{-\tfra{k+2}{2}}             \nonumber \\
           && \qquad+\bosb{2}{q^{(\epsilon-1)(k+2)}z}{-1}+~
              \bosc{2}{q^{(\epsilon-1)(k+1)-1}z}{0}  \Bigr\}: ,
\label{splitjm}\\
   J^{+}_{\rho}(z)&=& :\exp\left\{
         -\bosb{2}{q^{-k-2}z}{1}~
         -\bosc{2}{q^{-k-2+\rho}z}{0}  \right\}: ,
\label{splitjp}\\
   S_{\delta}(z)&=&:  \exp \Bigl\{
      - \bosa{k+2}{q^{-2}z}{-\tfra{k+2}{2}} \nonumber \\
     &&\qquad - \bosb{2}{q^{-k-2}z}{-1}
      - \bosc{2}{q^{-k-2+\delta}z}{0}
       \Bigr\} : ,
\label{splits}
%
\end{eqnarray}
In (\ref{splitjm}), $\pbosae{q^{-2}z}{-\frac{k+2}{2}}$ denotes the field
$$ \pbosae{q^{-2}z}{-\tfra{k+2}{2}}
    =\epsilon \Bigl\{
           (q-q^{-1})\sum_{n=1}^{\infty}\aPhi{\epsilon n}z^{-\epsilon n}~
         q^{(2\epsilon-{\tfra{k+2}{2}})n}+\tilde\aPhi{0}\log q \Bigr\}.$$
The factors $f_{\Phi}^{\lambda_l}(\{\epsilon\})$
and $f_{\phi\chi}(\{\epsilon\},\{\delta\},\{\rho\})$
are the contributions
from the $\Phi$ and the $\phi\chi$ sector, respectively:
\begin{eqnarray}
f_{\Phi}^{\lambda_l}(\{\epsilon\})&=&
    \Tr_{F^{\Phi}_l}\biggl(
        \zeta^{L^{\Phi}_0-\frac{c-2}{24}}y^{-\tilde \aPhi{0}}
        \prod_{i=1}^a e^{\pbosae{q^{-2}w_i}{-\frac{k+2}{2}}}
        \prod_{r=1}^m e^{\bosal{g}{2}{k+2}{q^k\zeta_r}{\frac{k+2}{2}}}\nonumber
\\
 &&\qquad \times \prod_{s=1}^n
e^{\bosal{1}{2}{k+2}{q^{k-1}z_s}{-\frac{k+2}{2}}}
        \prod_{j=1}^b e^{-\bosa{k+2}{q^{-2}t}{-\frac{k+2}{2}}}\biggr),
\label{fPhi}\\
f_{\phi\chi}(\{\epsilon\},\{\delta\},\{\rho\})
&=&\oint{dw_0\over 2\pi i w_0}\oint\frac{dw}{2\pi i}
   \Tr_{F^{\phi\chi}}\biggl(
        \zeta^{L^{\phi}_0+L_0^{\chi}-\frac{1}{12}}y^{-\tilde \aphi{0}}
        \prod_{i=1}^a e^{\bosb{2}{\tilde W^{\epsilon}_i}{-1}}
         \prod_{s=1}^n e^{\bosb{2}{Z_s}{0}}
        \nonumber\\
 &&\qquad \times
         \prod_{j=1}^b e^{-\bosb{2}{T_j}{-1}}
        \prod_{j=1}^b e^{-\bosb{2}{U_h}{1}}
         e^{\bosc{2}{ q^{-k-2}w}{0}}
         \prod_{i=1}^a e^{\bosc{2}{ W^{\epsilon}_i}{0}}
          \nonumber\\
 &&\qquad \times
         \prod_{s=1}^n e^{\bosc{2}{Z_s}{0}}
         e^{-\bosc{2}{ W_0}{0}}\prod_{j=1}^b e^{-\bosc{2}{T^{\delta}_j}{0}}
        \prod_{j=1}^b e^{-\bosc{2}{U^{\rho}_h}{0}}
\biggr).
\label{fphichi}
\end{eqnarray}
Here we have
set $T_j\equiv q^{-k-2}t_j, U_h\equiv q^{-k-2}u_h, W_0\equiv q^{-k-2}w_0,
\tdj\equiv q^{\delta_j}T_j, \urh\equiv q^{\rho_h}U_h,
Z_s\equiv q^{-2}z_s,
\wei\equiv q^{(\epsilon_i-1)(k+1)-1}w_i$, and  $\twei\equiv
q^{(\epsilon_i-1)(k+2)}w_i$.

      Now we evaluate the factors
$f_{\Phi}^{\lambda_l}(\{\epsilon\})$
and $f_{\phi\chi}(\{\epsilon\},\{\delta\},\{\rho\})$ separately.
    For the $\Phi$ sector, we obtain the following
result:

\begin{eqnarray}
f_{\Phi}^{\lambda_l}(\{\epsilon\})
&=&\eta(\tau)^{-1}\ \zeta^{\frac{l(l+2)+1}{4(k+2)}}
y^{-l}\ q^{l\sum\epsilon_i}\ \biggl[\fra{\prod ( q^k\zeta_r)^g\prod q^{k-2}z_s}
{\prod(q^{-2}t_j)^2}\biggr]^{l/2(k+2)}\nonumber \\
&&\quad\times \prod_{j<j'}G_S(t_j,t_{j'})\prod_{s<s'}G_{\Phi}(z_s,z_{s'})
\prod_{j,s}G_{\Phi S}(t_j,z_s)\nonumber \\
&&\quad\times \prod_{i<i'}G_{J^-}(w_i,w_{i'})\prod_{r<r'}
G_{\Psi}(\zeta_r,\zeta_{r'})
\prod_{j,i}G_{SJ^-}(t_j,w_i)\nonumber \\
&&\quad
\times \prod_{j,r}G_{S\Psi}(t_j,\zeta_r)\prod_{i,r}G_{J^-\Psi}(w_j,\zeta_{r})
\prod_{r,s}G_{\Psi\Phi}(\zeta_r,z_s)\prod_{i,s}G_{J^-\Phi}(w_j,z_{s}),
\end{eqnarray}
where the function $\eta(\tau)$ is
Dedekind's $\eta$-function
$$\eta(\tau)=\zeta^{1/24}\ (\zeta;\zeta)_{\infty}.$$
The expressions of the two point functions $G_S(t,t')$ etc., are
given in Appendix B.

    For the $\phi\chi$ sector, we have  a master formula ((56)
in Ref.\cite{Koa}).
Applying that formula to the case (\ref{fphichi}),
we obtain the  results.
\begin{eqnarray}
\lefteqn{f_{\phi\chi}(\{\epsilon\},\{\delta\},\{\rho\})
}\nonumber \\
&&=
\eta(\tau)\ y^{-1}\ q^{\sum\epsilon}\ ({q^{-2}\zeta};{\zeta})^{-a-b}_{\infty}
\ ({{q^2\zeta};{\zeta}})^{-c}_{\infty}\
({{\zeta};{\zeta}})^{-n}_{\infty}
\nonumber \\
&&
\qquad \times \oint{dw_0\over 2\pi i w_0}
\fra{\prod_1^b E(W_0,\tdj )\prod_1^cE(W_0,\urh )}
{\prod_1^aE(\wei ,W_0)\prod_1^nE(Z_s,W_0)}
\Prod{j,h}\fra{E(\tdj ,\urh)}{E(T_j,U_h)}
\Prod{i,h}\fra{E(\twei ,U_h)}{E(\wei ,\urh )}\nonumber \\
&&
\qquad\times
\Prod{i<i'}\fra{ E(\wei ,\weip)}{E_q(\twei ,\tweip;-2)}
\Prod{j<j'}\fra{E(\tdj ,\tdjp)}{E_q(T_j,T_{j'};-2)}
\Prod{h<h'}\fra{E(\urh , \urhp)}{E_q(U_h ,U_{h'};2)}\nonumber \\
&&
\qquad\times
\Prod{i,s}\fra{E(\wei ,Z_s)}{E_q(\twei,Z_s;-1)}
\Prod{i,j}\fra{E_q(\twei,\tdj;-2)}{E(\wei,\tdj)}
\Prod{i,j}\fra{E_q(Z_s,T_j;-1)}{E(Z_s,\tdj)}
\Prod{s,h}\fra{E_q(Z_s,U_h;1)}{E(Z_s ,\urh)}
\nonumber \\
&&
\qquad\times\
\fra{\prod_{i=1}^a
\vartheta_1(\wei+q^{-\sum\epsilon
-\sum\delta-\sum\rho}- W_0+y^2|\tau)
}{
\prod_{j=1}^b\vartheta_1(\tdj+q^{-\sum\epsilon
-\sum\delta-\sum\rho}- W_0+y^2
|\tau)} \nonumber \\
&&\qquad\times
\fra{\prod_{s=1}^n
\vartheta_1(Z_s+q^{-\sum\epsilon
-\sum\delta-\sum\rho}- W_0+y^2|\tau)
}{
\vartheta_1(q^{-\sum\epsilon
-\sum\delta-\sum\rho}+y^2|\tau)
\prod_{h=1}^c\vartheta_1(\urh+q^{-\sum\epsilon
-\sum\delta-\sum\rho}- W_0+y^2
|\tau)}.
\label{pctrace}
\end{eqnarray}
Here we
used the abridged notations $T_i$ to express ${\log T_i\over 2\pi i}$ etc.
in theta functions.
The functions $E(w,z)$ and $E_q(w,z;\alpha)$
are given by
\begin{eqnarray}
E(w,z)&=&{1\over \sqrt{z/w}}{\inp{z/w}{\zeta}
\inp{\zeta w/z}{\zeta}\over (\zeta;\zeta)^2_{\infty}},   \\
E_q(w,z;\alpha)&=&{1\over \sqrt{z/w}}\inp{q^{\alpha}z/w}{\zeta}
\inp{q^{\alpha}\zeta w/z}{\zeta}.
\end{eqnarray}

\section{ Residue Formula}

    In this section, we  analyze the  pole structures of the form factors.
We then derive the formulae giving the residues of corresponding poles.
The derivation given here is independent of the integral formulae obtained in
the
last section. In Appendix C,
we  give another derivation based on the integral formulae.

   Let us consider the following coefficients.
\begin{eqnarray}
&&
f^{\VEp_N\VEp_{N-1}..\VEp_2\VEp_1}_{\lambda\lambda_{N-1}
\lambda_{N-2}...\lambda_2\lambda_1\lambda}
(z_{N},z_{N-1},...,z_2,z_1) \nonumber
\\
&&\qquad=\Tr_{{\cal H}_{\lambda}}\Bigl(
q^{4\tilde L_0}q^{-J^3_0}
\cLl\Phi^{\lambda}_{\lambda_{N-1},\VEp_N}(z_{N})
\Phi^{\lambda_{N-1}}_{\lambda_{N-2},\VEp_{N-1}}(z_{N-1})
\cdots \Phi^{\lambda_2}_{\lambda_1,\VEp_2}(z_2)
\Phi^{\lambda_1}_{\lambda,\VEp_1}(z_1)
\Bigr).
\label{CNpff}
\end{eqnarray}

   Let us  investigate a local behavior of the product
$\Phi^{\lambda_2}_{\lambda_1,\VEp_2}(z_2)
\Phi^{\lambda_1}_{\lambda,\VEp_1}(z_1)$.
Since the $q$-VOs are intertwiner of $\uqa$, in the trace over IHWR,
 properties of any
matrix elements of the product are determined by those taken by
the highest weight states $\ket{\lambda}$s\cite{DFJMN}.
We therefore only have to investigate the pole structure of the two point
functions calculated in Sec.3.\par

   For example, let us consider the function (\ref{tfmmp}). Using the
definition of the Jackson integral (\ref{jackson}),
one can show that a simple pole at $q^{-2}z_1/z_2=1$ appears from the
first term in the
infinite series. One thus obtains the following local structures at the
neighborhood of the point $q^{-2}z_1/z_2=1$:
$$
\Phi^{\mu}_{\lambda_-,+}(z_2)\Phi^{\lambda_-}_{\lambda,-}(z_1)=
q^3(1-p){(q^2p;p)_{\infty}\over (q^{-2}p;p)_{\infty}}
{\xi(q^2;1,q^4)\over B_p(1-2ls,-2s)}
\delta^{\mu}_{\lambda}
{1 \over  1-q^{-2}z_1/z_2}{\rm id}+O(1).
$$

   In the same analysis for the remaining two point functions, one obtains
\begin{equation}
\Phi^{\mu}_{\lambda_{\pm},\VEp_2}(z_2)
\Phi^{\lambda_{\pm}}_{\lambda,\VEp_1}(z_1)
={\cal N}^{\lambda_{\pm}}_{\lambda}(\VEp_2,\VEp_1)
\delta_{\VEp_2,-\VEp_1}\delta^{\mu}_{\lambda}
{1\over 1-q^{-2}z_1/z_2}{\rm id}+O(1),
\label{gopetta}
\end{equation}
where
\begin{eqnarray}
{\cal N}^{\lambda_-}_{\lambda}(+,-)&=&
-q^4(1-p){\inp{q^2p}{p}\over \inp{q^{-2}p}{p}}
{\xi(q^2;1,q^4)\over B_p(1-2ls,-2s)}\nonumber\\
&=&-q{\cal N}_{\lambda_-}(\lambda;-,+),\\
{\cal N}^{\lambda_+}_{\lambda}(+,-)&=&
q(1-p){\inp{q^2p}{p}\over \inp{q^{-2}p}{p}}{\xi(q^2;1,q^4)\over
B_p(2(l+2)s,-2s)}\nonumber\\
&=&-q{\cal N}^{\lambda_+}_{\lambda}(-,+).
\end{eqnarray}

    Generally, in taking the trace $\Tr_{{\cal
H}_{\lambda}}\Bigl(\zeta^{L_0}\cdots
\Phi^{\mu}_{\lambda_{\pm},\VEp_2}(z_2)
\Phi^{\lambda_{\pm}}_{\lambda,\VEp_1}(z_1)\Bigr)$,
one must deal with additional poles. Namely, using the relation
\begin{equation}
\zeta^{L_0} \Phi^{\mu}_{\lambda,\VEp}(z)
 =  \Phi^{\mu}_{\lambda,\VEp}({\zeta z})\zeta^{L_0}
\end{equation}
and the cyclic property of trace,
one can obtain the trace $\Tr_{{\cal H}_{\rho}}\Bigl(\zeta^{L_0}\cdots
\Phi^{\lambda_{\pm}}_{\lambda,\VEp_1}(z_1)
\Phi^{\lambda}_{\rho,\VEp_2}(\zeta z_2)
\Bigr)$.
Therefore, one has to
take into acount the poles arising from
\begin{equation}
\Phi^{\sigma}_{\rho_{\pm},\VEp_1}(z_1)
\Phi^{\rho_{\pm}}_{\rho,\VEp_2}(\zeta z_2)
={\cal N}^{\rho_{\pm}}_{\rho}(\VEp_1,\VEp_2)
\delta_{\VEp_2,-\VEp_1}\delta^{\sigma}_{\rho}
{1\over 1-q^{-2}\zeta z_2/z_1}{\rm id}+O(1).
\label{gopettop}
\end{equation}

      In our case, $\zeta=q^4$.
One hence has to consider the both
contributions from (\ref{gopetta}) and (\ref{gopettop})
 in the calculation of  the residue at
$q^{-2}z_1/z_2=1$.

      Noting these facts,
as well as the commutation relation of the type II $q$-VOs
(\ref{crttvo}), we finally obtain the following residue
\begin{eqnarray}
\lefteqn{{{\rm res}\atop{q^{-2}z_1/ z_2=1}}
f^{\VEp_N\VEp_{N-1}...\VEp_1}_{\lambda\lambda_{N-1}...\lambda_1\lambda}
(z_N,z_{N-1},...,z_1)}\nonumber \\
&&=\delta^{\lambda_2}_{\lambda}\delta_{\VEp_2,-\VEp_1}
{\cal N}^{\lambda_1}_{\lambda}(-\VEp_1,\VEp_1)
 f^{\VEp_N\VEp_{N-1}...\VEp_3}_{\lambda\lambda_{N-1}...\lambda_3\lambda}
(z_N,z_{N-1},...,z_3)\nonumber \\
&&\qquad -{\cal N}^{\lambda}_{\lambda_1}(\VEp_1,-\VEp_1)\
\delta_{\tau_{N-2},-\VEp_1}R^{\rho_N\tau_{N-2}}_{\VEp_N\tau_{N-3}}\Bigl({z_N\over z_2}\Bigr)
R^{\rho_{N-1}\tau_{N-3}}_{\VEp_{N-1}\tau_{N-4}}\Bigl({z_{N-1}\over z_2}\Bigr)
\cdots R^{\rho_4 \tau_2}_{\VEp_3\tau_{1}}\Bigl({z_4\over z_2}\Bigr)
R^{\rho_3\tau_1}_{\VEp_3\VEp_{2}}\Bigl({z_3\over z_2}\Bigr) \nonumber \\
&&\times\sum_{\mu_1,...,\mu_{N-3}}
{ W}\left(\matrix{\mu_{N-3}&\lambda_{N-1}\cr
                            \lambda_{1}&\lambda \cr}{\Biggl |}{z_N\over
z_2}\right)
{ W}\left(\matrix{\mu_{N-4}&\lambda_{N-2}\cr
                            \mu_{N-3}&\lambda_{N-1} \cr}{\Biggl |}{z_N\over
z_2}\right)
\cdots
{ W}\left(\matrix{\mu_{1}&\lambda_{3}\cr
                            \mu_{2}&\lambda_4 \cr}{\Biggl |}{z_4\over
z_2}\right)
{ W}\left(\matrix{\lambda_{1}&\lambda_{2}\cr
                            \mu_{1}&\lambda_3 \cr}{\Biggl |}{z_3\over
z_2}\right)
\nonumber \\
&&\qquad\times
 f^{\rho_N\rho_{N-1}...\rho_3}_{\lambda_1\mu_{N-3}...\mu_1\lambda_1}
(z_N,z_{N-1},...,z_3).
\label{residue}
\end{eqnarray}

      This result satisfies all the characteristic features of the
Smirnov's third axiom. Namely,  the form factor has only a simple pole,
and the residue of the $N$-particle form factor is given by the $N-2$-particle
form factors as well as  by the $R$-matrix elements.
In addition to these features, our result shows that the form factor
in the higher spin XXZ model requires  face type Boltzmann weights
${ W}\left(\matrix{\lambda&\mu\cr
        \mu'&\nu \cr}{\Biggl |}{z}\right)$.

\section{Conclusion}
We have obtained  a free field realization of the type II $q$-vertex operators
of $\uqa$ of arbitrary  level $k$. Determining the cycles for the
Jackson integral associated with the screening charges,
we have showed that our type II $q$-vertex operators satisfies  the
$R$-matrix symmetry. We have also derived an integral formulae
for from factors in the spin $k/2$ XXZ model. It has been  shown that the
result has a proper analytic structure and
satisfies  the higher spin extension of the lattice analog of the Smirnov's
third axiom. Combining this result with those obtained in Ref.\cite{IIJMNT},
we now recover all the data necessary to  define
the  Riemann-Hilbert problem for form factors.

      Since our  form factor  is constructed
fully based on the representation theory of $\uqa$,
it is likely that the form factors in the massive integrable
quantum field theory ( e.g. higher spin
extension of the $SU(2)$ invariant Thirring model, which is
expected to be a certain scaling limit of our lattice model ) can  also
be understood in terms of the representation theory of
quantum algebra.
For this purpose, to investigate
the scaling limit of the form factors is an interesting problem\cite{Na}.

     On the other hand, recently the idea of ``vertex operator construction
of form factors'' has been  applied by Lukyanov directly to the $SU(2)$
invariant
Thirring model as well as the sine-Gordon model\cite{Lu}.
There, the Fadeev-Zamolodchikov generators, which correspond to our
type II $q$-vertex operators, were realized by means of free fields.
However, the expected underlying infinite dimensional symmetry, i.e. Yangian,
and the
representation theoretical meaning of his
vertex operators, i.e. intertwining property, are still unclear.
To make a  connection between Lukyanov's results with those obtained
in Ref.\cite{DFJMN,JMMN} and this paper could be helpful to clarify  these
points.

      We hope to discuss these subject in the future.
\section*{Acknowledgement}
The author would like to thank K.Iohara, M.Jimbo, M.Kashiwara,
 A. Matsuo, T.Miwa  and  A. Nakayashiki
for valuable discussions and suggestions. He is also grateful to RIMS,
where the most parts of this work were done,
for kind hospitality. This work was supported by Soryushi Shogakukai.

\appendix
\section{Operator Product Expansion Formulae}

We here summarize  the operator product expansion formulae among
the current $J^{+}(z)$, the screening operator ${S}(z)$ and the
$q$-vertex operator $\Phi_{l,0}(z)$.

    By splitting the operators $J^+(u)$ and $S(t)$
as (\ref{splitjp}) and (\ref{splits}), we obtain as follows.
\begin{equation}
   \begin{array}{rcll}
      \jpup(u) \Phi_{l,0}(z) &=&
	 q^{-l}\fra{ u - q^{k+l} z}{u-q^{k-l} z}
	 : \jpup(u) \Phi_{l,0}(z) :
	 & \quad |u|>q^{k-l}|z|
      \hh \\
      \jpdn(u) \Phi_{l,0}(z) &=&
	 q^{l}
	 : \jpdn(z) \Phi_{l.0}(z) :
	 & \quad
      \hh \\
      \Phi_{l,0}(z) \jpup(u) &=&
	 : \Phi_{l,0}(z) \jpup(u) :
	 & \quad
      \hh \\
      \Phi_{l,0}(z) \jpdn(u) &=&
	 \fra{z - q^{l-k} u}{z - q^{-l-k} u}
	 : \Phi_{l,0}(w) \jpdn(z) :
	 & \quad |z|>q^{-l-k}|u|
   \end{array}
\end{equation}
\begin{equation}
   \begin{array}{rcll}
      \jsup(t) \jpup(u) &=&
	 q
	 : \jpup(u) \jsup(t) :
	 & \quad
      \hh \\
      \jsup(t) \jpdn(u) &=&
	 q \fra{ t - q^{-2} u}{t - u }
	 : \jdn(u) \jsup(t) :
	 & \quad |t| > q^{-2} |u|
      \hh \\
      \jsdn(t) \jpup(u) &=&
	 q^{-1}\fra{ t - q^{2} z}{t-u}
	 : \jpup(u) \jsdn(t) :
	 & \quad |t| > |u|
      \hh \\
      \jsdn(t) \jpdn(u) &=&
	 q^{-1}
	 : \jpdn(u) \jsdn(t) :
	 & \quad
      \hh
   \end{array}
\end{equation}
\begin{equation}
   \begin{array}{rcll}
      \jpup(u) \jsup(t) &=&
	 q
	 : \jpup(u) \jsup(t) :
	 & \quad
      \hh \\
      \jpup(u) \jsdn(t) &=&
	 q\fra{ u- q^{-2} t}{u-t}
	 : \jpup(u) \jsdn(t) :
	 & \quad |u| > q^{-2} |t|
      \hh \\
      \jpdn(u) \jsup(t)  &=&
	 q^{-1}\fra{u - q^{2} t}{u-t}
	 : \jpdn(u) \jsup(t) :
	 & \quad |u| >  |t|
      \hh \\
      \jpdn(u) \jsdn(t) &=&
	 q^{-1}
	 : \jpdn(u) \jsdn(t) :
	 & \quad
   \end{array}
\end{equation}
\begin{equation}
   \begin{array}{rcll}
      \jsup(t) \Phi_{1,0}(z) &=&
	 ( q^{-2} t )^{-1/k+2}q^{-1}
         \fra{\inp{q^{k+1}z/t}{p}}{\inp{q^{k-1}z/t}{p}}
         : \jsup(t) \Phi_{1,0}(z) :
	 & \quad |t| > q^{k-1} |z|
      \hh \\
      \jsdn(t) \Phi_{1,0}(z) &=&
	   ( q^{-2} t )^{-1/k+2}q
	\fra{\inp{q^{k+1}pz/t}{p}}{\inp{q^{k-1}pz/t}{p}}
	 : \jsdn(t) \Phi_{1,0}(z) :
	 & \quad |t| > q^{k-1} p |z|
      \hh \\
      \Phi_{1,0}(z) \jsup(t) &=&
	 ( q^{k-2} z )^{-1/k+2}
	 \fra{\inp{q^{-k+1}p t/z}{p}}{\inp{q^{-k-1}p t/z}{p}}
	: \jsup(z) \Phi_{1,0}(z) :
	 & \quad |z| > q^{-k-1}p |t|
      \hh \\
      \Phi_{1,0}(z) \jsdn(t) &=&
	 ( q^{k-2} z )^{-1/k+2}
	 \fra{\inp{q^{-k+1} t/z}{p}}{\inp{q^{-k-1} t/z}{p}}
	 : \jsdn(t) \Phi_{1,0}(z) :
	 & \quad |z| > q^{-k-1} |t|
   \end{array}
\end{equation}
\begin{equation}
  \Phi_{l,0}(z)  \Phi_{l,0}(w) =
  (q^{k-2}z)^{l^{2}/2(k+2)}
   \fra{\indp{q^{2(1-l)}w/z}{p}{q^{4}}
    \indp{q^{2(1+l)}w/z}{p}{q^{4}}}
   {\indp{q^{2}w/z}{p}{q^{4}}^{2}}
  : \Phi_{l,0}(z)  \Phi_{l,0}(w) :
\end{equation}
Especially for $l=1$, we have
\begin{equation}
  \Phi_{1,0}(z)  \Phi_{1,0}(w) =\xi(w/z;1,q^4)
  : \Phi_{1,0}(z)  \Phi_{1,0}(w) :.
\end{equation}

\section{Two Point Functions}

We here give the expressions only for $
G_{S}(t,u), G_{S\Phi}(t,z), G_{\Phi S}(z,t)
$ and $G_{\Phi}(z,w)$, which are used in the text.
\begin{eqnarray}
G_S(t,u)&=&\cN{S}\ (q^{-2}t)^{2/k+2}
\fra{\indp{q^{-2}u/t}{p}{\zeta}\indp{q^{-2}\zeta t/u}{p}{\zeta}}
{\indp{q^{2}u/t}{p}{\zeta}\indp{q^{2}\zeta t/u}{p}{\zeta}},
\\
\cN{S}&=&\fra{\indp{q^{-2}\zeta}{p}{\zeta}}{\indp {q^{2}\zeta}{p}{\zeta}},
\end{eqnarray}

\begin{eqnarray}
G_{S\Phi}(t,z)&=&(q^{-2}t)^{-1/k+2}
\fra{\indp{q^{k+1}z/t}{p}{\zeta}\indp{q^{-k+1}\zeta t/z}{p}{\zeta}}
{\indp{q^{k-1}u/t}{p}{\zeta}\indp{q^{-k-1}\zeta t/u}{p}{\zeta}},
\\
G_{\Phi S}(z,t)&=&(q^{k-2}z)^{-1/k+2}
\fra{\indp{q^{-k+1}t/z}{p}{\zeta}\indp{q^{k+1}\zeta z/t}{p}{\zeta}}
{\indp{q^{-k-1}t/z}{p}{\zeta}\indp{q^{k-1}\zeta z/t}{p}{\zeta}},
\label{gps}
\end{eqnarray}

\begin{eqnarray}
G_{\Phi}(z,w)&=&\cN{\Phi}^2 \ (q^{k-2}z)^{1/2(k+2)}
\fra{\intp{q^{4}w/z}{q^4}{p}{\zeta}\intp{w/z}{q^4}{p}{\zeta}}
{\intp{q^{2}w/z}{q^4}{p}{\zeta}}\nonumber
\\
&&\qquad\qquad\times
\fra{\intp{q^{4}\zeta z/w}{q^4}{p}{\zeta}\intp{\zeta z/w}{q^4}{p}{\zeta}}
{\intp{q^{2}\zeta z/w}{q^4}{p}{\zeta}},
\\
\cN{\Phi}&=&
\fra{\intp{q^{4}\zeta }{q^4}{p}{\zeta}\intp{\zeta }{q^4}{p}{\zeta}}
{\intp{q^{2}\zeta}{q^4}{p}{\zeta}},
\end{eqnarray}

\section{Another Derivation of Residue Formula}
We here give another derivation of the residue formula (5.9) based on the
integral formulae for the form factors obtained in Sec.4.

      Let us consider the coefficients
\begin{eqnarray}
\lefteqn{
f^{++--}_{\lambda\lambda_+\lambda\lambda_+\lambda}
(z_4,z_3,z_2,z_1)
} \nonumber \\
&&=\oint\fra{du_1}{2\pi i}
\oint\fra{du_2}{2\pi i}
 <\Pdn{z_4}\Pup{z_3}[\Pdn{z_2},J^+(u_1)]_q[\Pup{z_1},J^+(u_2)]_q
>_{\lambda},
\label{formf}
\end{eqnarray}
where  we used the abridged notation
\begin{equation}
 <{\cal O}>_{\lambda}=\Tr_{{\cal H}_{\lambda}}\Bigl(
q^{4L_0}q^{-J^3_0}\cLl {\cal O}\Bigr).
\end{equation}

    The integrand in (\ref{formf}) consists of the following four terms.
\begin{eqnarray}
&&\qquad\quad<\Pdn{z_4}\Pup{z_3}\Pdn{z_2}J^+(u_1)\Pup{z_1}J^+(u_2)>_{\lambda},
\label{ffa}\\
&&-q<\Pdn{z_4}\Pup{z_3}J^+(u_1)\Pdn{z_2}\Pup{z_1}J^+(u_2)>_{\lambda},
\label{ffb}\\
&&-q<\Pdn{z_4}\Pup{z_3}\Pdn{z_2}J^+(u_1)J^+(u_2)\Pup{z_1}>_{\lambda},
\label{ffc}\\
&&q^2<\Pdn{z_4}\Pup{z_3}J^+(u_1)\Pdn{z_2}J^+(u_2)\Pup{z_1}>_{\lambda}.
\label{ffd}
\end{eqnarray}

        Let us investigate a pole structure at the point $q^{-2}z_2/z_3=1$.
There are
two sources of the poles. The first one is the correlation between the vertex
$\Phi_{1,0}(z)$ and the screening operator $S(t)$. The second one is
 the contour
integral with respect to the argument of $J^{+}(u)$.

    We first consider the poles arising from the product
$\Pup{z_3}J^+(u_1)\Pdn{z_2}$
in (\ref{ffb}) and (\ref{ffd}).
Let us consider  the poles from the first source.
Using (\ref{gps}),(\ref{pctrace}) and
(\ref{svodn}) for $\Phi^{\lambda}_{\lambda_+,+}(z_2)$,
the trace
of the product $\Pp{z_3}\Pp{z_2}S_{\delta}(t)$
contributes the factors:
\begin{equation}
G_{\Phi S}(z_3,t)G_{\Phi S}(z_2,t)\label{pscorr}
\label{ggtrace}
\end{equation}
in the $\Phi$ sector, and
\begin{equation}
\fra{E_q(Z_3,T;-1)E_q(Z_2,T;-1)}{E(Z_3,T^{\delta})E(Z_2,T^{\delta})}
\label{ppstrace}
\end{equation}
in the $\phi\chi$ sector.
After scaling $t\to q^{k-1}z_2 t$,
we find  that a simple pole at $q^{-2}z_2/z_3=1, t=1$
arising from $G_{\Phi S}(z_3,t)$ cancelled by a zero at $t=1$ in
$G_{\Phi S}(z_2,t)$ as well as, in (\ref{ppstrace}) with $\delta=1$,
 a simple pole at $t=1$
cancelled by a zero at $q^{-2}z_2/z_3=1, t=1$. As a result, we have no
poles arising from the first source in (\ref{ffb}) and (\ref{ffd}).

     Next let us consider the poles from the second source
in (\ref{ffb}) and (\ref{ffd}).
Let us first consider the integral with respect to $u_1$.
The poles in the $u_1$-plane arise from the correlations among
$J^+(u_1), \Pp{z_2}, \Pp{z_3}$ and $S(t)$. By using  the formula
(\ref{pctrace}) ,
we find the trace of
the product $\Pp{z_3}J^+(u_1)\Pp{z_2}$ contributes  the factor
\begin{equation}
\fra{E_q(Z_3,U;1)E_q(U,Z_2;1)}{E(Z_3,U^{\rho})E(U^{\rho},Z_2)}
=\left\{\matrix{
\fra{\inp{q^{k+5}z_3/u}{q^4}\inp{q^{k+1}z_2/u}{q^4}}
{\inp{q^{k+3}z_3/u}{q^4}\inp{q^{k-1}z_2/u}{q^4}}
&{\rm for}\quad\rho=1\cr
&\cr
\fra{\inp{q^{-k+1}u/z_3}{q^4}\inp{q^{-k+5}u/z_2}{q^4}}
{\inp{q^{-k-1}u/z_3}{q^4}\inp{q^{-k+3}u/z_2}{q^4}}
&{\rm for}\quad\rho=-1\cr
}\right.
\end{equation}
whereas the trace of  $J^+(u_1)S(t)$ has a contribution by
\begin{equation}
\fra{E(U^{\rho},T^{\delta})}{E(U,T)}=
\left\{
\matrix{
1&{\rm for}\quad \delta=\rho=\pm1\cr
&\cr
{q^{-1}}\ \fra{\inp{q^{k+1}z_2t/u}{q^4}\inp{q^{-k+3}u/z_2t}{q^4}}
{\inp{q^{k-1}z_2t/u}{q^4}\inp{q^{-k+5}u/z_2t}{q^4}}
&{\rm for}\quad\delta=1,\quad\rho=-1\cr
&\cr
{q}\ \fra{\inp{q^{k-3}z_2t/u}{q^4}\inp{q^{-k+7}u/z_2t}{q^4}}
{\inp{q^{k-1}z_2t/u}{q^4}\inp{q^{-k+5}u/z_2t}{q^4}}
&{\rm for}\quad\delta=-1,\quad\rho=1\cr
}\right. .
\end{equation}
We hence find that, only in the case $\delta=1,\quad \rho=-1$,
the contour is pinched at $q^{-2}z_2/z_3=1 ,\quad t=1$
by the simple poles at $u_1=q^{k-1}z_2$
inside of the contour and the one at $u_1=q^{k+1}z_3$ outside of the contour.
This implies the  simple pole at $q^{-2}z_2/z_3=1$ when $t=1$
 in (\ref{ffb}) and (\ref{ffd}).

   Next let us consider the contour integral with respect to $u_2$.
For this purpose, we manipulate (\ref{ffb}) to obtain
\begin{eqnarray}
&&-q^2\fra{1-q^{k-1}z_1/u_2}{1-q^{k+1}z_1/u_2 }r\left(\fra{z_4}{z_3}\right)
r\left(\fra{z_1}{q^4z_3}\right)\sum_{\mu,\nu}
W^+_\mu\left(\fra{z_4}{z_3}\right)
W^{\mu}_\nu\left(\fra{z_1}{q^4z_3}\right)\nonumber \\
&&\qquad\qquad\times
<\Phi^{\mu}_{\lambda,+}(z_4)J^+(u_1)
\Phi^{\lambda}_{\lambda_+,+}(z_2)J^+(u_2)\Phi^{\lambda_+}_{\nu,+}(q^4z_3)
\Phi^{\nu}_{\mu,+}(z_1)>_\mu,
\label{manua}
\end{eqnarray}
 where
$W_{\mu}^+(z)\equiv
W\left(\matrix{\lambda&\lambda_{+}\cr
                            \mu&\lambda \cr}{\Biggl |}z\right)$ and
$W_{\nu}^{\mu}(z)\equiv
W\left(\matrix{\mu&\lambda\cr
                            \nu&\lambda_+ \cr}{\Biggl |}z\right)$.
Here we used the $R$-matrix symmetry (\ref{crttvo}),
the commutation relation (4.1), the cyclic property
of trace and
the following relations.
\begin{eqnarray}
J^+(u)\Pp{z} &=& q^{-1} \fra{1-q^{k+1}z/u}{1-q^{k-1}z/u} \Pp{z}J^+(u), \\
\qint{ \Pp{z} , J^+(u) }_q &=& -q^{k}z/u \fra{1-q^2}{1-q^{k+1}z/u}J^+(u)\Pp{z}.
\end{eqnarray}

For (\ref{ffd}), the analogous manipulation leads to
\begin{eqnarray}
&&r\left(\fra{z_4}{z_3}\right)
r\left(\fra{z_1}{q^4z_3}\right)\sum_{\mu,\nu}
W^+_\mu\left(\fra{z_4}{z_3}\right)
W^\mu_\nu\left(\fra{z_1}{q^4z_3}\right)\nonumber \\
&&\qquad\qquad\times
<\Phi^{\mu}_{\lambda,+}(z_4)J^+(u_1)
\Phi^{\lambda}_{\lambda_+,+}(z_2)J^+(u_2)\Phi^{\lambda_+}_{\nu,+}(q^4z_3)
\Phi^{\nu}_{\mu,+}(z_1)>_\mu,
\label{manub}
\end{eqnarray}

     Using (\ref{svoup}) for $\Phi^{\lambda}_{\lambda_+,+}(z_2)$,
the same analysis for the product
$\Phi^{\lambda}_{\lambda_+,+}(z_2)J^+(u_2)\Phi^{\lambda_+}_{\nu,+}(q^4z_3)$
as the above implies that, only in the case
$\nu=\lambda$ with  $\delta=-1,\quad \rho=1$,
 pinching of the contour with respect to $u_2$ occurs and
a simple pole at $q^{-2}z_2/z_3=1$ arises.

    We now calculate the corresponding residues.
Remarkably, the following operator identities are verified.
\begin{equation}
:\Pp{z_3}J^+_{-1}(u)\Pp{z_2}S_{1}(q^{k-1}z_2t):=1
\end{equation}
at  $t=1,\quad q^{-2}z_2/z_3=1$ and $u=q^{k-1}z_2$, and
\begin{equation}
:\Pp{z_2}J^+_{-1}(u)\Pp{q^4z_3}S_{1}(q^{k+3}z_3t):=1
\end{equation}
at  $t=1,\quad q^{-2}z_2/z_3=1$ and $u=q^{k+3}z_3$.

\begin{equation}
:S_{-1}(q^{k+1}z_3t)\Pp{z_3}J^+_{1}(u)\Pp{z_2}:=1
\end{equation}
at  $t=1,\quad q^{-2}z_2/z_3=1$ and $u=q^{k-1}z_2$, and
\begin{equation}
:S_{-1}(q^{k+1}z_2t)\Pp{z_2}J^+_{1}(u)\Pp{q^4z_3}:=1
\end{equation}
at  $t=1,\quad q^{-2}z_2/z_3=1$ and $u=q^{k+3}z_3$.

     Due to these identities, we obtain the residues of the terms (\ref{ffb})
 and (\ref{ffd}) as follows.
\begin{eqnarray}
\lefteqn{{\cal N}^{\lambda}_{\lambda_+}(+,-)\oint\fra{du}{2\pi i}
<\Pdn{z_4}\Pup{z_1}J^+(u)>_{\lambda}
}\nonumber \\
&&-q^2{\cal N}^{\lambda_+}_{\lambda}(-,+)\fra{1-z_1/q^4z_3}{1-q^{2}z_1/q^4z_3}
r\left(\fra{z_4}{z_3}\right)
r\left(\fra{z_1}{q^4z_3}\right)\sum_{\mu=\lambda_+,\lambda_-}
W^+_{\mu}\left(\fra{z_4}{z_3}\right)
W^{\mu}_{\lambda}\left(\fra{z_1}{q^4z_3}\right)\nonumber \\
&&\qquad\qquad\times \oint\fra{du}{2\pi i}
<\Phi^{\mu}_{\lambda,+}(z_4)J^+(u)\Phi^{\lambda}_{\mu,+}(z_1)
>_{\mu},
\label{resb}
\end{eqnarray}
for (\ref{ffb}) and
\begin{eqnarray}
\lefteqn{{\cal N}^{\lambda}_{\lambda_+}(+,-)q^2\oint\fra{du}{2\pi i}
<\Pdn{z_4}J^+(u)\Pup{z_1}>_{\lambda}
}\nonumber \\
&&
+q^2{\cal N}^{\lambda_+}_{\lambda}(-,+)
 r\left(\fra{z_4}{z_3}\right)
r\left(\fra{z_1}{q^4z_3}\right)\sum_{\mu=\lambda_+,\lambda_-}
W^+_{\mu}\left(\fra{z_4}{z_3}\right)
W^{\mu}_{\lambda}\left(\fra{z_1}{q^4z_3}\right)\nonumber \\
&&\qquad\qquad\times \oint\fra{du}{2\pi i}
<\Phi^{\mu}_{\lambda,+}(z_4)J^+(u)\Phi^{\lambda}_{\mu,+}(z_1)
>_{\mu},
\end{eqnarray}
for (\ref{ffd}).

     For the remaining terms  (\ref{ffa}) and (\ref{ffc}),
we make analogous manipulation to (\ref{manua}) and (\ref{manub}):
\begin{eqnarray}
\lefteqn{
q^{-2}\fra{1-q^{k-1}z_1/u_2}{1-q^{k+1}z_1/u_2}
\fra{1-q^{k+5}z_3/u_2}{1-q^{k+3}z_3/u_2}
r\left(\fra{z_4}{z_3}\right)
r\left(\fra{z_1}{q^4z_3}\right)\sum_{\mu,\nu}
W^+_\nu\left(\fra{z_4}{z_3}\right)
W^\mu_\lambda\left(\fra{z_1}{q^4z_3}\right)
}\nonumber \\
&&\qquad\times <\Phi^{\mu}_{\lambda,+}(z_4)
\Phi^{\lambda}_{\lambda_+,+}(z_2)J^+(u_1)\Phi^{\lambda_+}_{\nu,+}(q^4z_3)J^+(u_2)
\Phi^{\nu}_{\mu,+}(z_1)>_\mu
\label{ffa1}
\end{eqnarray}
and
\begin{eqnarray}
\lefteqn{q^{2}\fra{1-q^{k-1}z_1/u_2}{1-q^{k+1}z_1/u_2}
\fra{1-q^{k-1}z_2/u_1}{1-q^{k+1}z_2/u_1}
r\left(\fra{z_4}{z_3}\right)
r\left(\fra{z_1}{q^4z_3}\right)\sum_{\mu,\nu}
W^+_\nu\left(\fra{z_4}{z_3}\right)
W^\mu_\nu\left(\fra{z_1}{q^4z_3}\right)
}\nonumber \\
&&\qquad\times
<\Phi^{\mu}_{\lambda,+}(z_4)J^+(u_1)
\Phi^{\lambda}_{\lambda_+,+}(z_2)J^+(u_2)\Phi^{\lambda_+}_{\nu,+}(q^4z_3)
\Phi^{\nu}_{\mu,+}(z_1)>_\mu
\label{ffa2}
\end{eqnarray}
for (\ref{ffa}), whereas
\begin{eqnarray}
\lefteqn{
-\fra{1-q^{k+5}z_3/u_2}{1-q^{k+3}z_3/u_2}
r\left(\fra{z_4}{z_3}\right)
r\left(\fra{z_1}{q^4z_3}\right)\sum_{\mu,\nu}
W^+_\mu\left(\fra{z_4}{z_3}\right)
W^\mu_\nu\left(\fra{z_1}{q^4z_3}\right)
}\nonumber \\
&&\qquad\times
<\Phi^{\mu}_{\lambda,+}(z_4)
\Phi^{\lambda}_{\lambda_+,+}(z_2)J^+(u_1)\Phi^{\lambda_+}_{\nu,+}(q^4z_3)J^+(u_2)
\Phi^{\nu}_{\mu,+}(z_1)>_\mu
\label{ffc1}
\end{eqnarray}
and
\begin{eqnarray}
\lefteqn{
-q^{2}\fra{1-q^{k-1}z_2/u_1}{1-q^{k+1}z_2/u_1}
r\left(\fra{z_4}{z_3}\right)
r\left(\fra{z_1}{q^4z_3}\right)\sum_{\mu,\nu}
W^+_\nu\left(\fra{z_4}{z_3}\right)
W^+_\mu\left(\fra{z_1}{q^4z_3}\right)
}\nonumber \\
&&\qquad\times
<\Phi^{\mu}_{\lambda,+}(z_4)J^+(u_1)
\Phi^{\lambda}_{\lambda_+,+}(z_2)J^+(u_2)\Phi^{\lambda_+}_{\nu,+}(q^4z_3)
\Phi^{\nu}_{\mu,+}(z_1)>_\mu
\label{ffc2}
\end{eqnarray}
for (\ref{ffc}).

     By the same analysis as before, we
find  only a simple pole at $q^{-1}z_2/z_3=1$
in (\ref{ffa1})$\sim$(\ref{ffc2}), and we  obtain the following residues.
\begin{eqnarray}
\lefteqn{
{\cal N}^{\lambda_+}_{\lambda}(-,+)\left\{
\fra{1-q^{k+5}z_3/u}{1-q^{k+3}z_3/u}
\fra{1-q^{k-1}z_1/u}{1-q^{k+1}z_1/u}
+q^2\fra{1-q^{k-1}z_2/u}{1-q^{k+1}z_2/u}
\fra{1-z_1/q^4z_3}{1-q^{2}z_1/q^4z_3}
\right\}
r\left(\fra{z_4}{z_3}\right)
r\left(\fra{z_1}{q^4z_3}\right)}\nonumber \\
&&\qquad\times
\sum_{\mu=\lambda_+,\lambda_-}
W^+_{\mu}\left(\fra{z_4}{z_3}\right)
W^\mu_{\lambda}\left(\fra{z_1}{q^4z_3}\right) \oint\fra{du}{2\pi i}
<\Phi^{\mu}_{\lambda,+}(z_4)J^+(u)\Phi^{\lambda}_{\mu,+}(z_1)
>_{\mu}
\end{eqnarray}
 from
(\ref{ffa}) and
\begin{eqnarray}
\lefteqn{
{\cal N}^{\lambda_+}_{\lambda}(-,+)\left\{
-\fra{1-q^{k+5}z_3/u}{1-q^{k+3}z_3/u}
-q^2\fra{1-q^{k-1}z_2/u}{1-q^{k+1}z_2/u}
\right\}
r\left(\fra{z_4}{z_3}\right)
r\left(\fra{z_1}{q^4z_3}\right)
}\nonumber \\
&&\qquad\times
\sum_{\mu=\lambda_+,\lambda_-}
W^+_{\mu}\left(\fra{z_4}{z_3}\right)
W^\mu_{\lambda}\left(\fra{z_1}{q^4z_3}\right)\oint\fra{du}{2\pi i}
<\Phi^{\mu}_{\lambda,+}(z_4)J^+(u)\Phi^{\lambda}_{\mu,+}(z_1)
>_{\mu}
\label{resc}
\end{eqnarray}
 from
(\ref{ffc}).

      Combining the results (\ref{resb})$\sim$(\ref{resc}), we
finally obtain the
residue
 \begin{eqnarray}
\lefteqn{
{{\rm res}\atop{{ q^{-2}z_2/z_3=1}}}
f^{++--}_{\lambda\lambda_+\lambda\lambda_+\lambda}
(z_4,z_3,z_2,z_1)
}\nonumber \\
&&={\cal N}^{\lambda}_{\lambda_+}(+,-)
<\Pdn{z_4}\Pupm{z_1}>_{\lambda}\nonumber \\
&&\qquad-q{\cal N}^{\lambda_+}_{\lambda}(-,+)
\fra{1-z_1/q^4z_3}{1-q^{2}z_1/q^4z_3}r\left(\fra{z_4}{z_3}\right)
r\left(\fra{z_1}{q^4z_3}\right)\sum_{\mu=\lambda_+,\lambda_-}
W^+_{\mu}\left(\fra{z_4}{z_3}\right)
W^\mu_{\lambda}\left(\fra{z_1}{q^4z_3}\right)\nonumber \\
&&\qquad\qquad \times
<\Phi^{\mu}_{\lambda,+}(z_4)\Phi^{\lambda}_{\mu,-}(z_1)>_{\mu}
\label{restt}
\end{eqnarray}

     From (\ref{restt}),  the residue at $q^{-2}z_1/z_2=1$ is obtained as
follows.
Moving  the vertex $\Pupm{z_1}$
and $\Pdnm{z_1}$ in the both sides to the left
by using the trace property and renaming the arguments as
$z_4\to z_3, z_3\to z_2, z_2\to z_1$ and $z_1\to q^4 z_4$,
we obtain
\begin{eqnarray}
\lefteqn{{{\rm res}\atop{{ q^{-2}z_1/z_2=1}}}
f^{-++-}_{\lambda_+\lambda\lambda_+\lambda\lambda_+}(z_4,z_3,z_2,z_1)
}\nonumber \\
&&={\cal N}^{\lambda}_{\lambda_+}(+,-)
f^{+-}_{\lambda_+\lambda\lambda_+}({z_4},{z_3})\nonumber \\
&&\qquad
-{\cal N}^{\lambda_+}_{\lambda}(-,+)R^{-+}_{-+}\left(\fra{z_4}{z_2}\right)
R^{++}_{++}\left(\fra{z_3}{z_2}\right)\sum_{\mu=\lambda_+,\lambda_-}
W^\mu_{\lambda}\left(\fra{z_4}{z_2}\right)
W^+_{\mu}\left(\fra{z_3}{z_2}\right)
f^{-+}_{\lambda\mu\lambda}(z_4,z_3).
\label{residuef}
\end{eqnarray}
This is the special case of the general result (5.9).
The general formula can be obtained from (\ref{residuef})
by using the $R$-matrix symmetry.

\end{document}